\documentclass[]{aa}
\usepackage{psfig} 
\newcommand{\be}{\begin{equation}}
\newcommand{\en}{\end{equation}}

\def\zabs{$z_{\rm abs}$}
\def\zem{$z_{\rm em}$}
\def\lya{Ly$\alpha$ }

\def\h2{H$_2$}

\def\kms{km~s$^{-1}$}
\def\dela{$\Delta\alpha/\alpha$~}
\begin{document}
%
%\maketitle
\title{{Probing the cosmological variation of the fine-structure constant:
Results based on VLT-UVES sample}
\thanks{
Based on observations collected
at the European Southern Observatory (ESO), under the Large Programme
"QSO Absorption Line Systems" ID No. 166.A-0106 with UVES on the 8.2m KUEYEN
telescope operated at the Paranal Observatory, Chile}
}
\titlerunning{fine-structure constant}  
\author{Hum Chand\inst{1}, Raghunathan Srianand\inst{1},  Patrick Petitjean\inst{2,3} \& Bastien Aracil\inst{2}}
\institute{
IUCAA, Post Bag 4, Ganeshkhind, Pune 411 007, India\\
%~\email{anand@iucaa.ernet.in}\\ 
Institut d'Astrophysique de Paris -- CNRS, 98bis Boulevard 
Arago, F-75014 Paris, France\\
LERMA, Observatoire de Paris, 61 Rue de l'Observatoire,
   F-75014 Paris, France\\
}
\date{Received date/ Accepted date}
\offprints{R.Srianand}
%\maketitle
%\markboth{}{}
%\pagerange{\pageref{firstpage}--\pageref{lastpage}}
%\label{firstpage}

\abstract{
Development of fundamental physics relies 
on the constancy of various fundamental quantities such as
the fine structure constant. Detecting or
constraining the possible time variations of these fundamental
physical quantities is an important step toward a complete
understanding of basic physics. High quality absorption lines
seen in the spectra of distant QSOs allow one to probe time
variations of several of these quantities.
Here we present the results from a detailed many-multiplet 
analysis, {to detect the possible variation of fine-structure constant},
performed using high signal-to-noise ratio, ($\sim$70 per pixel), high 
spectral resolution ($R\ge$45000)
observations of 23 Mg~{\sc ii} systems detected toward 18 QSOs in the redshift 
range $0.4\le z\le 2.3$ obtained using UVES at the VLT.
We validate our procedure and define the selection criteria
that will avoid possible systematics using detail analysis
of simulated data set. The spectra of Mg~{\sc ii} doublets
and Fe~{\sc ii} multiplets are generated considering
variations in $\alpha$ and specifications identical to
that of our UVES spectra. We show our Voigt profile
fitting code recovers the variation in $\alpha$ very
accurately when we use single component systems and
multiple component systems that are not heavily blended.
Spurious detections are frequently seen when we use
heavily blended systems or the systems with very weak
lines. Thus we avoided heavily blended systems and the
systems with Fe~{\sc ii} column density $<2\times10^{12}$
cm$^{-2}$ while analysing the UVES data.
To make the analysis transparent and
accessible to the community for critical scrutiny all the 
steps involved in the analysis are presented in detail.
The weighted mean value of the variation in ${\bf \alpha}$ 
obtained from our analysis over the redshift 
range ${ 0.4\le z\le 2.3}$ is 
${ \Delta\alpha/\alpha}$~=~${ (-0.06\pm0.06)\times10^{-5}}$. 
The median redshift of our sample is 1.55 and corresponds to
a look-back time of 9.7 Gyr in the most favored cosmological model today.
The 3$\sigma$ upper limit on the time variation of $\alpha$
is ${ -2.5\times 10^{-16} ~{\rm yr}^{-1}\le(\Delta\alpha/\alpha\Delta t) 
\le+1.2\times 10^{-16}~{\rm yr}^{-1}}$. 
To our knowledge this
is the strongest constraint from quasar absorption line
studies till date.
}
\maketitle

\section{Introduction}
Contemporary theories of fundamental interactions, such as SUSY GUT or
Super-string theories, treating gravity and quantum mechanics in a
consistent way, not only predict a dependence of fundamental physical
constants with energy (which has been observed in high energy
experiments)
but allow for their cosmological time and space
variations(Uzan, 2003).  Detecting or
constraining the possible time variations of fundamental
physical quantities is an important step toward a complete
understanding of basic physics.
In the framework of standard Big-bang models one can  probe the 
evolution of various physical quantities over the elapse time using 
measurements that are performed at different redshifts ($z$). 
As the energy of atomic transitions depend on the electromagnetic 
coupling constant $\alpha$ [its well measured 
laboratory value is $e^2/\hbar c$=1/137.03599958(52)
(see {Mohr \& Taylor 2000})], its possible time variation  will be 
registered in the form of small shifts in the absorption line spectra 
seen toward high-z QSOs 
(Bahcall, Sargent \& Schmidt 1967). Initial attempts 
to measure the variation in $\alpha$ were based on alkali-doublets 
(Wolfe, Brown \& Roberts 1976; Levshakov 1994; and Potekhin \& 
Varshalovich 1994; Cowie \& Songaila, 1995; Murphy et al. 2001). 
This method (AD method) uses the difference in the wavelengths 
of the doublets originating from the same ground state
(i.e., $^2$S$_{1/2}\rightarrow^2$P$_{3/2}$ and $^2$S$_{1/2}\rightarrow^2$P$_{1/2}$ 
transitions).  The constraints on the variation in $\alpha$,
\dela, is obtained by assuming that the measured difference in the wavelength
centroid of the doublets is proportional to $\alpha^2$ 
to the lowest order. 
The best constraint obtained using 
this method is $\Delta\alpha/\alpha$ = $(-0.5\pm1.3)\times10^{-5}$
(Murphy et al. 2001).

Like absorption doublets one can also use the central wavelength of
multiple emission lines originating from the same initial excited 
state. For example, Bahcall et al. (2003) use the nebular O~{\sc iii} emission
lines at $\lambda5007$ and $\lambda4959$ originating 
from $^1$D$_2$ excited level and derive
\dela = $(-2.0\pm1.2)\times10^{-4}$ based on 73 quasar
spectra over the redshift range 0.16$\le$z$\le$0.70. 

Studies based on  molecular absorption lines seen in the radio/mm 
wavelength range are more sensitive than that based on optical/UV
absorption lines.  Till now good constraints are available 
for two systems, 
\dela = $(-0.10\pm0.22)\times 10^{-5}$ at $z$ = 0.2467 and 
\dela = $(-0.08\pm0.27)\times10^{-5}$ at $z$ = 0.6847 if one
assumes constant proton g-factor (Murphy et al. 2001a). 
It is also possible to have joint constraints on the 
fundamental constants using molecular lines (see
Chengalur \& Kanekar, 2003).
However, such studies at high $z$ are not available due to
the lack of molecular absorption systems at high $z$.

The most suitable and accurate method for measuring \dela 
at high redshift is
called the many-multiplet method (MM {method}). This is a generalization 
of the AD method and
was introduced by Dzuba et al. (1999). Unlike the AD method, this method uses 
absolute wavelength measurements of 
numerous absorption lines from different species. 
This method has been shown to provide an order of magnitude improvement 
in the measurement of \dela compared to AD method (Murphy et al. 2003 and 
references there in). However, in the bargain, this is the most 
vulnerable method against a number of systematics. 

Accurate laboratory wavelengths (up to an accuracy of few m\AA) are 
available for most 
of the important
transitions that are regularly detected in QSO spectra (see Table.~1).
The sensitivity coefficients of the different line transitions from different 
multiplets to the variation in $\alpha$ were
computed using many-body calculations taking into account the 
dominant relativistic effects (Dzuba et al. 1999 \& 2002). If one
uses the parametrization of these authors then the accuracy to probe
the possible variations in $\alpha$ depends very much on how well
one measures the wavelength of the absorption lines at high $z$.

In simple terms, MM method exploits the fact that the energy of 
different line transitions varies differently for a given change 
in $\alpha$. For example, rest wavelengths of Mg~{\sc ii} doublets 
and Mg~{\sc i} are fairly insensitive to small changes in 
$\alpha$ thereby providing good anchors for measuring the 
systemic redshift. Whereas the rest wavelengths 
of Fe~{\sc ii} multiplets are very sensitive to small variations 
in $\alpha$. Thus constraining  relative shifts between an 
anchor and different Fe~{\sc ii} lines allows to study the
variations in $\alpha$. However the  accuracy 
depends on how well the absorption line profiles are modeled. 

The absorption profiles are usually modeled using
multiple Voigt profiles that are defined by column density ($N$), velocity
dispersion($b$) and redshift in addition to the rest-wavelength of the
species. In general the Voigt profile decomposition is robust.
However, the profile decomposition is not unique when the absorption
lines are heavily blended and/or if the signal-to-noise ratio in
the spectrum is not high enough. In addition, in real data, small relative 
shifts between absorption lines from different species 
can be introduced due to various systematic effects such as 
chemical and ionization inhomogeneities in the absorbing region, 
incorrect wavelength calibration, isotopic abundances, 
and atmospheric dispersion effects etc. It is usually argued 
that these random systematic effects can be canceled by 
using large number of measurements. 
Indeed, applying the method to 143 systems over a large redshift range 
0.2~$<$~$z$~$<$~3.7, Murphy et al (2003) measured 
\dela = ${(-0.54\pm0.12)}\times10^{-5}$.

The detailed work based on MM method has been performed by only
one group using simultaneous fitting of absorption line profiles 
of species covering a wide range of ionization states. 
Their important results need to be confirmed independently
using a data set that is optimized for the purpose.  This forms
the main motivation of this work. One of the main criticism 
made to the
MM method is the apparent lack of transparency in the complex
analysis (Bahcall et al. 2003).
Here, we perform similar analysis
to the one used by Webb and collaborators on a homogeneous and very
high quality data set.  To make the whole process accessible
for critical scrutiny, we describe in detail the procedure,
and the individual fits. The main idea is that it is better
to derive strong constraints on a small, although statistically
significant, sample of well chosen systems rather than 
loose constraints on a larger sample of strongly blended systems.

The details of the observations and the 
quality of the data used in the analysis are discussed in 
Section 2. In Section 3, we discuss simulations that are performed
to validate our Voigt profile fitting code in extracting the
variations in $\alpha$. We also use the simulation results to
construct an optimum selection criteria that will ensure a best 
possible detection limit in \dela. In Section 4, we list all
the Mg~{\sc ii}, Fe~{\sc ii} systems that are selected in our
sample. Comments on individual Voigt profile fitting are given 
in Section 5. Discussion and conclusions
are presented in Section 6.

 \begin{figure*}
\centerline{\vbox{
\psfig{figure=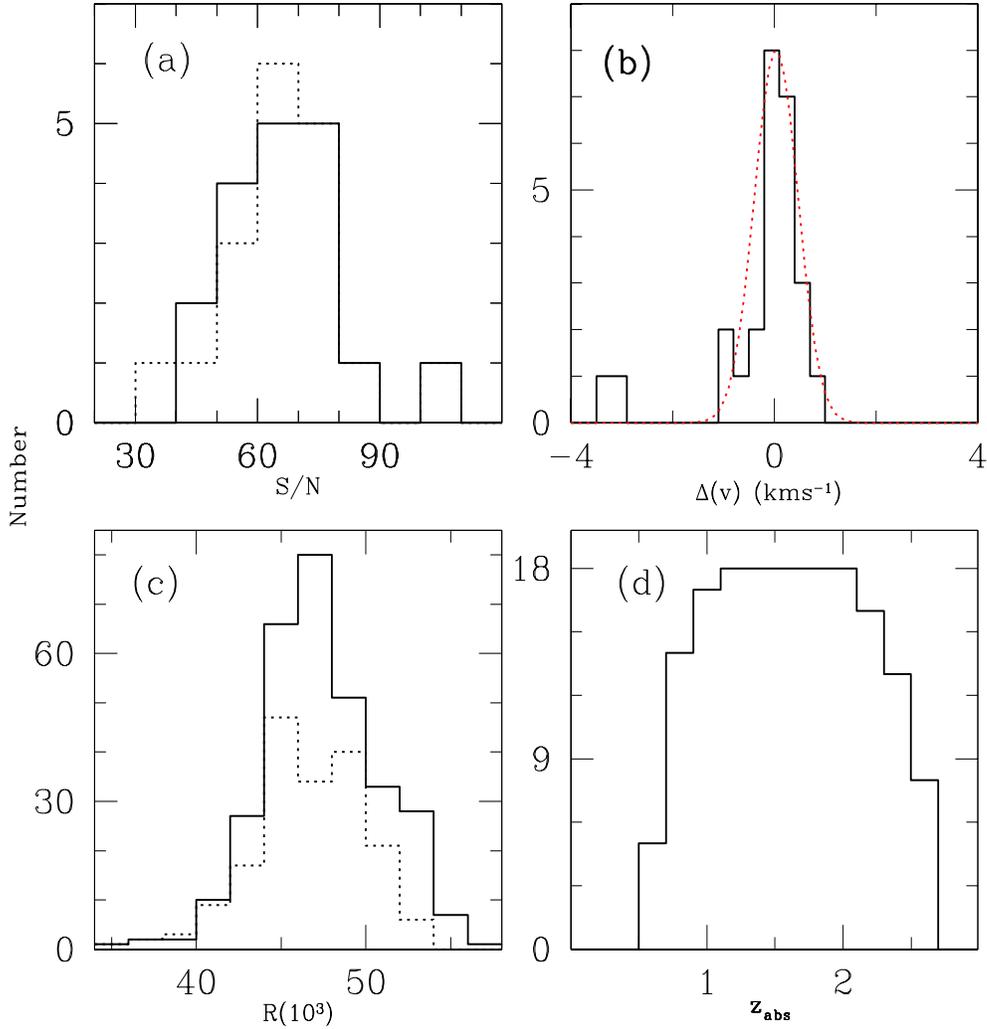,height=15.cm,width=14.cm,angle=0.}
%comph1.ps
}}
\caption[]{{\bf Properties of our sample:} 
Panel(a) shows the distribution of
median signal-to-noise (S/N) ratio per pixel computed over 30\AA~
intervals. Typical S/N = 60$-$80 per pixel is achieved over the 
wavelength range 4000$-$8000 \AA. Panel (b) shows the 
distribution of measured 
relative velocity shifts between Fe~{\sc ii}$\lambda$2344 and 
Fe~{\sc ii}$\lambda$2600 lines in our sample. 
As the q coefficients are similar for 
these transitions the velocity shift is expected to be distributed 
around zero in the case of good wavelength calibrations. Thus
the plot provides an internal consistency check for the wavelength
calibration. We notice that the accuracy in the wavelength 
calibration is of the 
order of 1-3 m\AA~ in the wavelength range of interest.
Panel (c) shows the distribution spectral resolution measured from 
the calibration spectrum. 
 Typical spectral resolution 
achieved is larger than 44,000. Continuous and dotted distributions in these 
panels are for the blue (4000-5500\AA) and red (5500-8500\AA) 
spectral range respectively. Panel (d) gives the histogram of the number 
of sight lines at a given redshift in which Mg~{\sc ii} and Fe~{\sc ii} 
absorption lines are in the observed wavelength range. All the 18 sight lines 
in our sample cover the redshift range between 1 and 2 uniformly.
}
\label{figobs}
\end{figure*}
 
\begin{figure*}
%\centerline{\vbox{
\psfig{figure=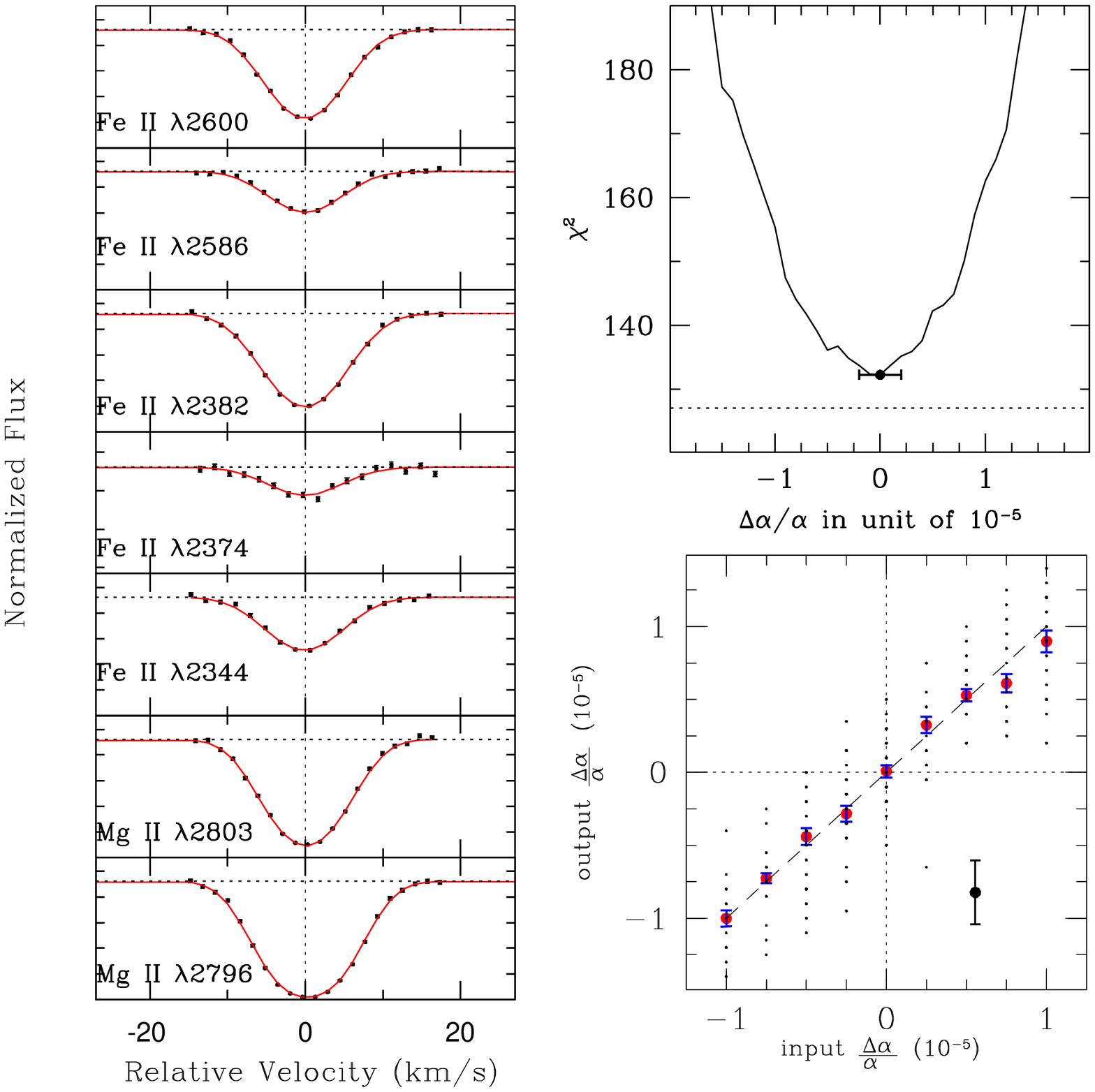,height=15.cm,width=18.0cm,angle=360}
%}}
\caption[]{ {\bf Illustration of the methodology and reliability of
our Voigt profile fitting code:} The panels in the left hand-side
show the velocity plot of the simulated Mg~{\sc ii} and Fe~{\sc ii}
absorption lines (points with error-bars) and the best fitted Voigt
profiles (solid curve). The panel in the top right hand-side 
shows the $\chi^2$ as a function of \dela. The best fitted value
and $1\sigma$ error is shown in the panel. This is consistent with
the input value of \dela = 0. Lower right panel shows the
relationship between the input and recovered values of \dela. 
Each point is a result obtained from a single realization (there are
30 of them for a given input \dela). Typical errors in these
measurements are shown in the bottom right corner of the panel.
The dark circles with error bars are the weighted mean value
of \dela obtained from 30 realizations. 
}
\label{figsing}
\end{figure*}

\section{Observations and data sample}

The data used in this study have been obtained using the 
Ultra-violet and Visible 
Echelle Spectrograph (UVES) mounted on the ESO KUEYEN 8.2~m telescope 
at the Paranal observatory for the ESO-VLT  Large Programme 
``QSO absorption lines". This corresponds to a homogeneous sample of 18 QSO
lines of sight suitable for studying various properties of the 
inter-galactic medium over a  redshift  range 1.7$-$3.2. 
All the objects were observed in good seeing conditions with
1 arcsec slit width. The data were reduced using the UVES 
pipeline, a set of procedures implemented in a dedicated context 
of MIDAS, the ESO data reduction package. The main characteristics of 
the pipeline is to perform a precise inter-order background subtraction 
for science frames 
and master flat-fields, and to allow for an optimal extraction of the object 
signal rejecting cosmic ray impacts and performing sky-subtraction at the 
same time. Usually a given wavelength range is covered by more than 
6 exposures and removal of cosmic rays and odd flux fluctuations in 
a given exposure is robust.
The reduction is checked step by step. Wavelengths are corrected 
to vacuum-heliocentric values and individual 1D spectra are combined 
together. Air-vacuum conversions and heliocentric corrections were done using
standard conversion equations (Elden (1966), {Stumpff}(1980)). The slit was always oriented 
along the parallactic angle and calibration exposures were taken
before or after the scientific exposures. Thus most of the important 
concerns raised by Murphy et al., are minimised in our observations.
As we are using $\chi^2$ minimization method to detect
sub pixel scale shifts the error spectrum is very crucial for our
analysis. Great care was taken in computing the error spectrum
while combining the individual exposures. Our final error is the
quadratic sum of appropriately interpolated weighted mean errors
and possible errors due to scatter in the individual flux
measurements.

In all the cases, spectra cover the wavelength range of 3000-10000\AA. 
A first guess continuum is fitted
with an automatic continuum fitting code. However for the system that
are eventually used in the analysis, a local continuum normalization
is performed using lower order polynomials. In all cases we have used
the flux in the core of the saturated lines to estimate the background
subtraction uncertainties. Corrections to this are applied 
whenever it is needed. We notice that continuum fitting is not 
unique in the \lya forest region and in the IR region where atmospheric 
lines are plenty.  Therefore  we have avoided
the absorption lines that are falling
in the lower wavelength side of \lya emission from the quasar and 
lines that could be blended with atmospheric lines in our
analysis.                          

The properties of the spectra in our sample are summarized in
Fig.~\ref{figobs}. Typical S/N$\sim60-80$ per pixel (typically 0.035\AA~
wide) is achieved in the whole wavelength range of interest (see left
hand top panel in Fig.~\ref{figobs}). This is
a factor 2 improvement compared to the one used earlier 
for the analysis of \dela (see Fig.~2 of Murphy et al. 2003). The spectral
resolution 
measured from the lamp spectrum is typically $\ge 44,000$. 

\begin{table*}[ht]
%\begin{center}
\caption{Summary of atomic parameters that are used in our analysis}
%
%First Column show the species used in our measurements.
%         Column 2 and 4 show respectively the rest wavelength of the species, with 
%         reference given by ID in column 3. The $q$ value in column 5 is taken from 
%         Dzuba et al. (2003). The available uncertainty are quoted in the bracket. Column 6
%         give the ID of species used in Table.3 to refer to the species used in measurements.
%          Column 7 and 8 give respectively the oscillator strength and corresponding reference. }
\begin{tabular}{llcllclc}
\hline\hline
\\
{Species} &{$\lambda_0$(\AA)} & Ref.($\lambda_0$) &{$\omega_{0}(cm^{-1})$} &{$q$} &id& $f$ & Ref.($f$) \\ 
\hline \\
Mg~{\sc i}  &2852.96310(8) & L  &35051.277(1)    &$+86(10)$      &i   &1.830 & A\\
Mg~{\sc ii} &2796.3543(2)  & L  &35760.848(2)    &$+211(10)$     &a   &0.6123  & A\\
            &2803.5315(2)  & L  &35669.298(2)    &$+120(10)$     &b   &0.3054  & A\\
Al~{\sc ii} &1670.7887(1)  & M  &59851.972(4)    &$+270(30)$     &j   &1.88   & B\\
Si~{\sc ii} &1526.70709(2) & M  &65500.4492(7)   &$+50(30)$      &k   &0.133  & A\\
            &1808.01301(1) & M  &55309.3365(4)   &$+531(30)$     &l   &0.00208  & A\\
%Ca~{\sc ii} &3934.777      & O  &25414.427       &$420^{*}$   &m   & 0.6346   & A\\
Fe~{\sc ii} &1608.45085(8) & P  &62171.625(3)    &$-1200(300)$ &c  & 0.0577  & C\\
            &2344.2130(1)  & Q  &42658.2404(2)   &$+1254(150)$   &d   & 0.114    & D\\
            &2374.4604(1)  & Q  &42114.8329(2)   &$+1640(150)$   &e   & 0.0313  & D\\
            &2382.7642(1)  & Q  &41968.0642(2)   &$+1498(150)$    &f   & 0.320   & D\\
            &2586.6496(1)  & Q  &38660.0494(2)   &$+1520(150)$   &g   & 0.0691  & D\\
            &2600.1725(1)  & Q  &38458.9871(2)   &$+1356(150)$   &h   & 0.239   & D\\
\hline
\multicolumn {8}{l}{Reference:}\\
\multicolumn {8}{l}{\emph{oscillator strength:-} (A) Morton D.C. (1991);
(B) Prochaska et al.(2001) (C) Morton D. C(2003);} \\
\multicolumn {8}{l}{(D) Bergeson et al. (1996);}\\
\multicolumn {8}{l}{\emph{wavelength:-} (L) Pickering et al. (1998);
(M) Griesmann U. \& Kling R.(2002); (P) Pickering et al. (2002)}\\
\multicolumn {8}{l}{(Q) Nave et al. (1991)}\\
%\multicolumn {8}{l}{ $*$ The dependence of $w$ on $\alpha$ for this line is given by $w=w_{0}+q_{1}x+q_{2}y$.}\\
%\multicolumn {8}{l}{where $x={\alpha_{z}/\alpha_{0}}^{2}-1$ and $y={\alpha_{z}/\alpha_{0}}^{4}-1$.The value in table}\\
%\multicolumn {8}{l}{refer to $q_{1}$, while its $q2$ value is 16$cm^{-1}$ (Dzuba V.A. et. al. 1999).}\\  
\end{tabular}
\label{tabat}
%\end{center}
\end{table*}

We have investigated the accuracy of wavelength
calibration using the lamp spectra that are extracted in the same
way as the object. We notice that individual emission lines are well modeled
by Gaussian. This confirms that the instrumental profile can be modeled
with single Gaussian while fitting the  Voigt profiles.
The median difference between the extracted and the
actual wavelengths in most of the settings are close to zero.
The root mean square of this deviation is always smaller than 3 m\AA 
in all the settings.
We confirm the robustness of the wavelength calibration using
other consistency checks. In particular, the wavelengths
of Fe~{\sc ii} transitions strongly depend on $\alpha$ but the relative
position of the lines ({apart from Fe~{\sc ii}$\lambda1607$}) does not depend much on $\alpha$. 
This means that, whatever the variation of
$\alpha$ is, the relative velocity shifts between Fe~{\sc ii}
transitions should be consistent with zero in 
the case of good calibration. This is precisely what is observed 
(Panel b in Fig.~\ref{figobs}). 
Note in addition that the Fe~{\sc ii} rest wavelengths are known to very high
accuracy ($\sim$0.1m\AA).
In principle similar exercise can be performed using Si~{\sc iv}
and C~{\sc iv} absorption lines.
However, unlike Fe~{\sc ii} lines, C~{\sc iv} wavelengths are known only 
to 2 m\AA~ accuracy. Even in this case the velocity shifts measured 
are consistent 
with zero within the allowed laboratory wavelength uncertainties
(Petitjean \& Aracil, 2003).
When fitting the narrow doublets and 
multiplets from a given species in individual settings, 
we have allowed for some adjustments in the
instrumental resolution. 
This seems to be an important step as we realized 
that the seeing was much better than the slit width during most of the 
observing run. The metal line systems in the reduced spectra are
identified using standard procedure.

The next step is to validate our Voigt profile fitting codes and
have some estimate of various possible systematic effects. 
In order to do this we first simulate the spectra with specifications
similar to a typical UVES spectrum and perform the analysis on them.
The details of the spectral simulations and the data analysis process
is described in the following section.

\section{Chasing the systematics using simulated data}

As one is trying to measure \dela through Voigt profile decomposition 
of absorption lines, various systematic effects can cause false 
alarm detection 
of \dela. The effects of random systematics 
can be averaged out by combining large number of measurements. 
However, the best strategy will be to have an a priori estimate of
various systematic effects so that a proper selection criteria
can be applied while choosing the sample to achieve the best
possible results. 
In this section we try to understand various effects using 
simulated spectra. Such an exercise is also important to see how well
our Voigt profile fitting code recovers the input value of 
\dela that is used while generating the spectrum.

\subsection{Generation of absorption lines}

%%%%%%%%%%%%%

In this section, we simulate data for different
assumed values of \dela and apply
our procedure to recover the input value of \dela.
We mainly concentrate on absorption 
lines from the Mg~{\sc ii} doublet and  Fe~{\sc ii} multiplets. 
We assume the ratio, $N$(Mg~{\sc ii})/$N$(Fe~{\sc ii}), to be constant 
in individual components. 
In order to mimic real data, the Voigt profile of the absorption line
is convolved with an instrumental profile and appropriate noise is added 
to the spectrum. We model the instrumental profile as a single Gaussian with 
$\sigma=1.69$ pixels (with pixel size of 0.03~\AA) appropriate for 
$\lambda/\Delta\lambda = 42,000$ as achieved in a typical UVES observations
(note however that the actual resolution achieved is somewhat higher
in the real data).

The noise in the spectrum is the combination of photon noise, readout noise,
and the residuals from background subtraction. While the former dominates 
the error in the continuum the latter two will be the source of error in 
the core of saturated lines. We notice from our UVES data
that the root mean square of residual  at the bottom of
highly saturated lines is $\simeq$0.4\% of the unabsorbed
continnum flux.
The readout noise added to a typical normalized spectrum, 
mimics a Gaussian distribution  
with zero mean and a standard deviation of $4\times10^{-3}$.
The photon noise spectrum is obtain using Poisson statistic.
To obtain a spectrum with a given 
signal-to-noise ratio, (S/N), we have  generated  
Poisson random numbers ($P[(S/N)^{2}]$), with mean equals to 
$(S/N)^{2}$ and then assigned 
photon noise  in the continuum of normalized spectrum, to be 
$(P[(S/N)^{2}]-(S/N)^{2})/(S/N)^{2}$, having Poisson distribution 
with $\sigma$ = (S/N)$^{-1}$. In the absorbed portion of the spectrum this
value is scaled by square root of the intensity. In all our simulations
we use the signal-to-noise in the continuum to be 70, corresponding to what 
is achieved in our data.

In order to simulate a spectrum incorporating a varying fine-structure constant 
($\alpha_{z}$), we use the analytic fitting function given by Dzuba et al. (2002),
\begin{equation}
\omega=\omega_{0}+qx.
\end{equation}
Here $\omega_{0}$ and $\omega$ are, respectively, the {vacuum} wave 
number (in the units of $cm^{-1}$) measured in the laboratory 
and in the absorption system at  redshift $z$, $x$ is a dimensionless number
defined by $ x=(\alpha_{z}/\alpha_{0})^{2}-1 $ where $\alpha_{0}$ 
refers to the present 
value of the fine-structure constant and  $\alpha_{z}$ its value at redshift $z$. 
The numerical value of the parameter $q$ that are obtained using 
many body relativistic calculations (see Dzuba et al. 1999, 2002) 
for different species are 
listed in the fifth column of the Table~\ref{tabat}. 
{Originaly $\omega$ is defined as $\omega=\omega_{0}+q_1x+q_2y$
with $y={(\alpha_{z}/\alpha_{0})}^{4}-1$ and $x$ as defined above(Dzuba et al., 1999).
When \dela is small q is basically ${q_1+2q_2}$.}
The table also gives the laboratory values of wavelengths 
({with standard isotopic composition for different elements})
($\lambda_0$), wave-numbers ($\omega_0$) and oscillator strengths ($f$) 
used in this study. References to various parameters used
are also provided in this table.

\subsection{Voigt profile fitting code}
\begin{figure}
%\centerline{\vbox{
\psfig{figure=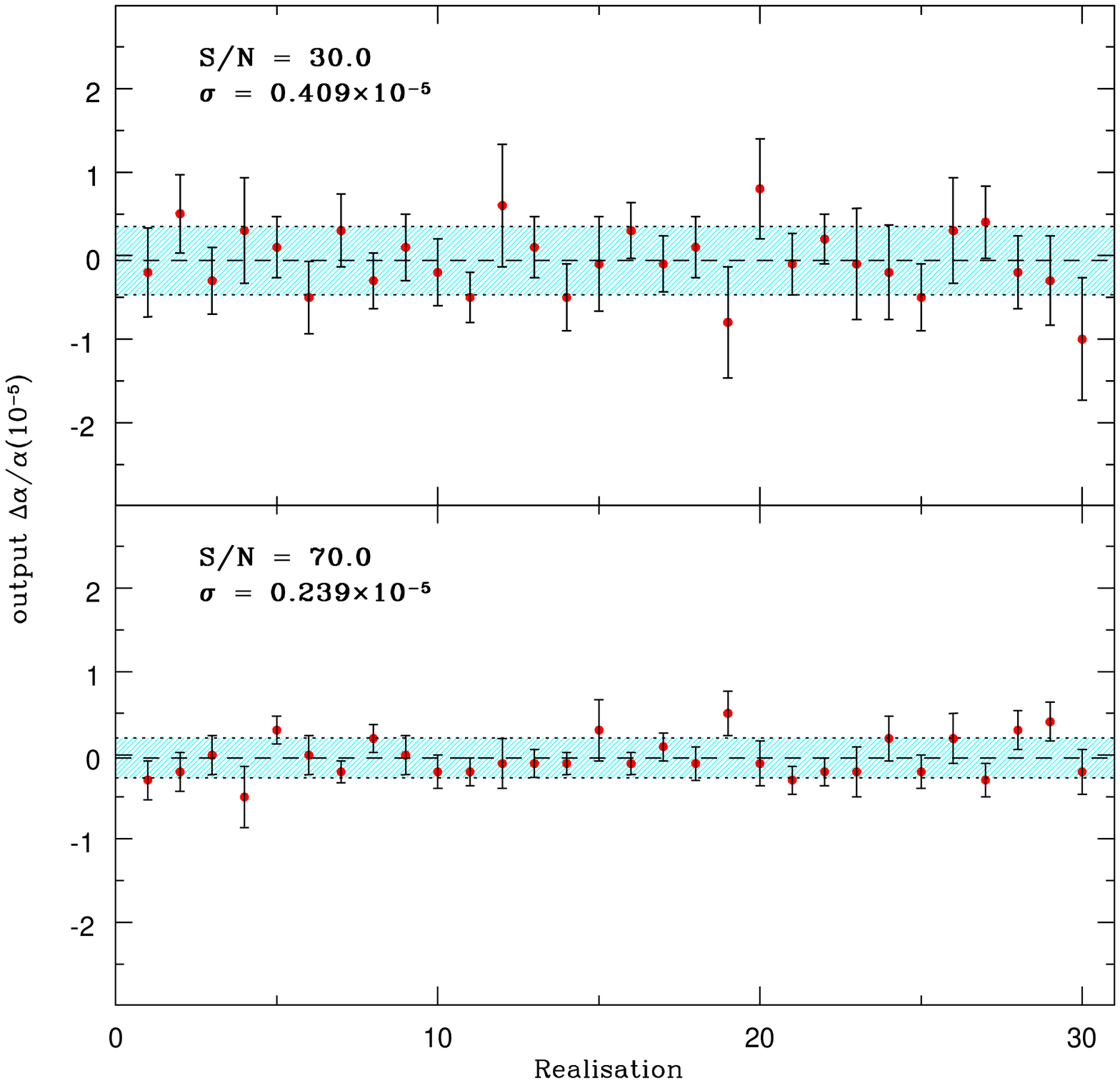,height=9.cm,width=8.cm,angle=00}
%}}
\caption{{\bf The effect of signal-to-noise ratio:}
The figure shows the measured
$\Delta\alpha/\alpha$ and $1\sigma$ error bar from 30 realizations of
absorption  systems all simulated with an input value $\Delta\alpha/\alpha=0$. 
The shaded region is the standard deviation (1$\sigma$) in the measured values 
around the mean value (shown by horizontal dashed line). The 
absorption systems are simulated using
random realizations of $N$, $b$ and noise spectrum. 
We obtained  $\sigma = 0.409\times10^{-5}$ and
0.239$\times10^{-5}$  based on 30 realizations with signal-to-noise
30 and 70 respectively.
It is apparent from the figure that better S/N will improve the 
measurements of $\Delta\alpha/\alpha$. 
}
\label{figsbyn}
\end{figure}

\begin{figure}
%\centerline{\vbox{
\psfig{figure=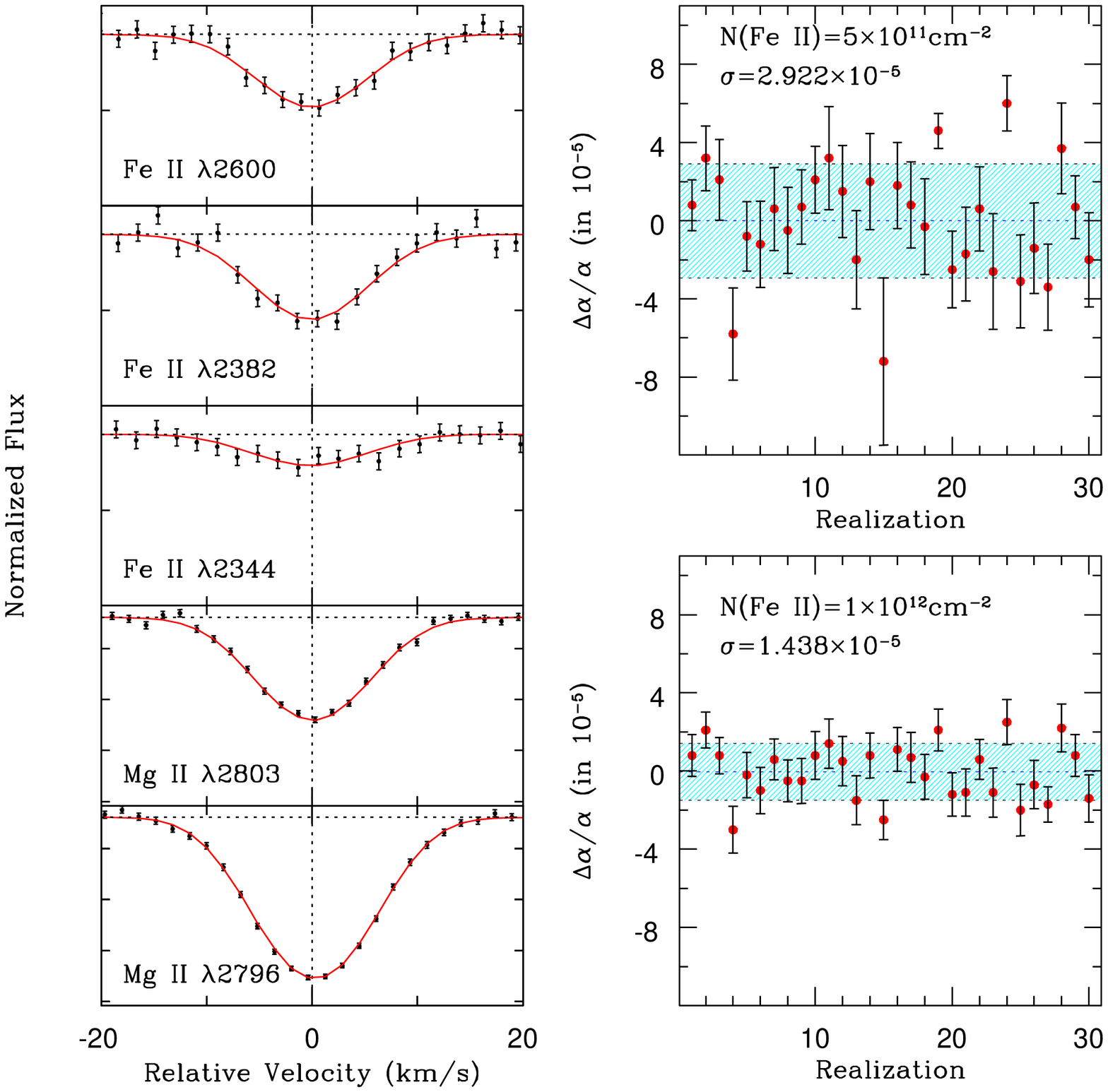,height=9.cm,width=8.cm,angle=00}
%}}
 \caption{{Simulations of weak lines:} The left hand side panels 
show the velocity plot of the absorption lines simulated in one of
the realizations. The best fitted Voigt profiles are overplotted. 
The right-hand side panels give the recovered value of \dela in individual
simulations for two values of $N$(Fe~{\sc ii}). In all cases the 
input \dela is zero and only the error spectrum is changed. 
The shaded regions in these panels gives the standard deviation
around the mean. The numerical values are also given in each
panel.
}
\label{figweak}
\end{figure}

To measure the relative shifts between an  anchor and  Fe~{\sc ii} lines,
we fit all the lines simultaneously. In addition, we have modified our 
Voigt profile fitting code (Khare et al. 1997) to take into account the dependence
of rest wavelengths of different species on $\alpha$. 
We improved the convolution algorithm 
using a 41 point Gaussian quadrature integration scheme.
lines in high signal-to-noise data. The FWHM of the Gaussian kernel
is considered to be  $\lambda_{\rm r}/R$. Here 
$\lambda_{\rm r}$ is the median wavelength over the region of absorption line 
being fitted and $R$ is the spectral resolution. We have taken  
a $\pm3\sigma$ ($\sigma=FWHM/2.3548$) wavelength
range around each pixel of absorption line for convolution. 
The Gaussian quadrature integration method is adopted because 
it provides the most accurate estimate of the area under the 
Gaussian kernel that is  used for normalizing the truncated
Gaussian probability function.
Our code also takes into account the variation of spectral 
resolution across the spectrum.
 
\subsection {Measurement of \dela in a single component systems}

In order to find the best \dela consistent with our data,
we fit the absorption lines using the modified
rest wavelengths (see equation (1)), varying \dela  in the range 
from $-5.0\times10^{-5}$ to 
$5.0\times10^{-5}$ in steps of $0.1\times10^{-5}$.
For each value we obtain $\chi^2_{min}$ by varying $N$, $b$ and $z$.
The value of \dela at which $\chi^2_{\rm min}$ is minimum is accepted as
the best possible \dela value, provided the reduced $\chi^2$
of fit is $\simeq 1$. Following the standard statistical
procedure (Press et al.(2000) see pages p690-691) 
we assign 1$\sigma$ errors  to the above defined best 
value of \dela by computing the required change in 
\dela so that $\Delta\chi^{2}= \chi^{2}-\chi^2_{\rm min}=1 $. To be on the
conservative side we have taken the maximum change in \dela around the
best value of \dela that is required to produce $\Delta\chi^{2}=1$ as our
error on \dela. 
 \begin{figure*}
%\centerline{\vbox{
\psfig{figure=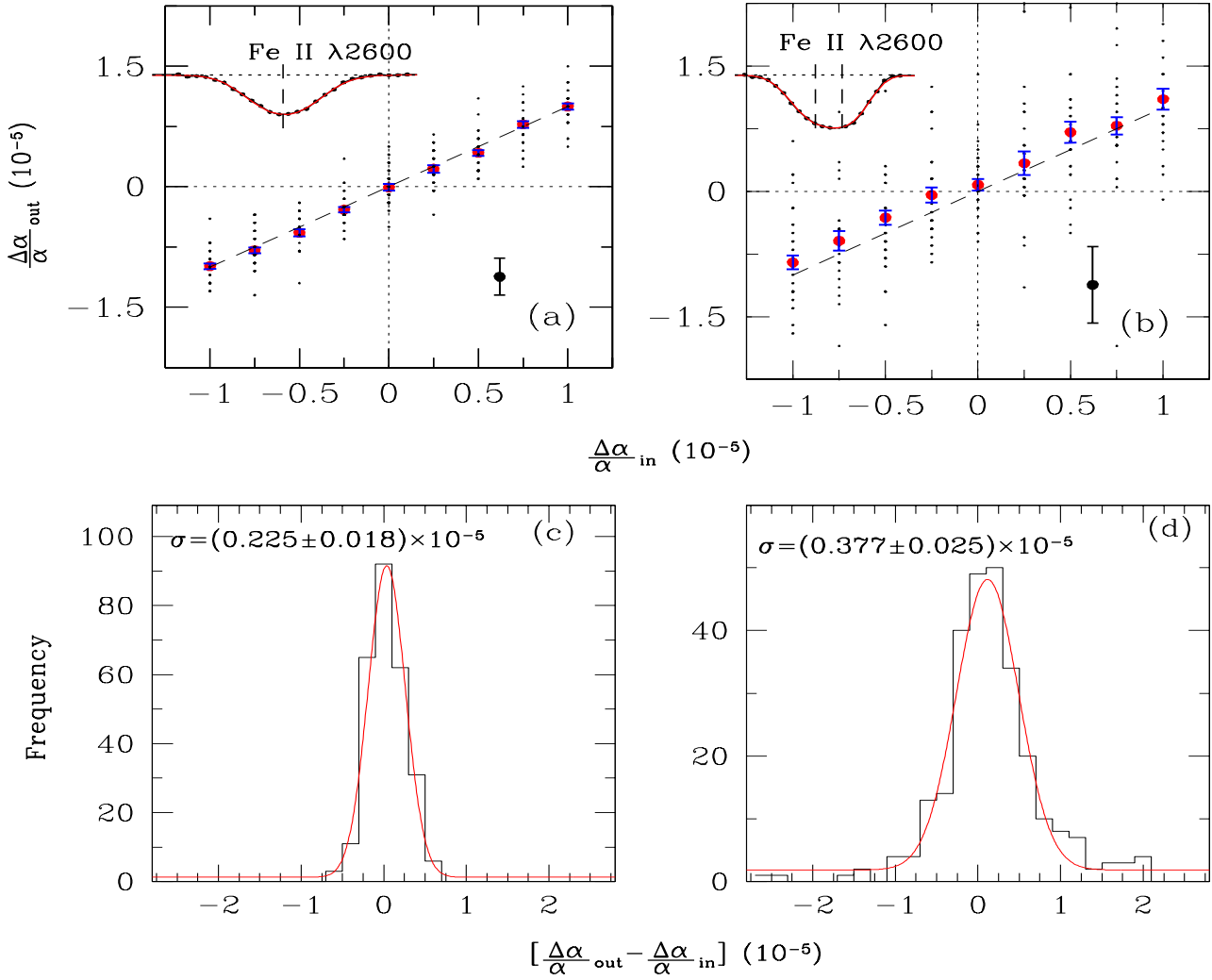,height=15.cm,width=18.0cm,angle=360}
%}}
\caption[]{ {\bf Effect of blending in the determination of \dela:}
Absorption spectra of Mg~{\sc ii} and Fe~{\sc ii} were
simulated for given $N$, $b$, and spectral resolution similar to 
that of our data, introducing spectral shifts corresponding to a given
value of \dela. Top panels show the relationship between
the input and derived value of $\Delta\alpha/\alpha$ in the case
of a single clean component (left-hand side) and a blend of two components
(right-hand side).
A typical absorption profile is also shown in these panels.
Dots are the values from individual
realizations; the points with the error bars are the weighted mean
over 30 realizations.
The lower panels give the distribution of the recovered $\Delta\alpha/\alpha$
around the true one. Single (left) and blended (right) cases are
considered respectively.
Fits by Gaussian distributions are over-plotted.
}
\label{figdou}
\end{figure*}

We test the method discussed above using a simulated 
single component system (see Fig~\ref{figsing}). The profiles of 
Mg~{\sc ii} and Fe~{\sc ii} 
absorption lines that are generated for \dela = 0 are shown in the left 
side column of the figure. In the top right side panel we show 
the best fitted $\chi^2$ as a function of \dela. The location
of the minimum and 1$\sigma$ error in \dela are shown using a point
with an error bar.
In order to investigate the systematics in detail, we have simulated 
single component
systems with column densities in the range 10$^{13}$ to 10$^{16}$
cm$^{-2}$ (we consider the number of systems with a 
given $N$(Mg~{\sc ii}) is proportional to $N$(Mg~{\sc ii})$^{-1.4}$ 
(Srianand \& Khare 1994)), velocity dispersion uniformly distributed
in the range 3 to 10 \kms and input \dela varying between 
$-10^{-5}$ and $+10^{-5}$.  In addition to that we have also varied the seed 
for the random number generator that produces the error spectrum. 
We use our Voigt profile fitting 
code to recover \dela in each case.  The recovered value of \dela in 
each realization is plotted (dots) as a function of the input 
\dela in the lower right panel in Fig.~\ref{figsing}. 
Typical error in an individual measurements is also shown 
in the plot. There are about
30 realizations for a given value of input \dela with a 
varied $N$, $b$ and noise spectrum. 
The dark circles with error bars are the  weighted mean value of 
the output \dela using all the 30 measurements. The weighted 
mean recovered values of \dela are consistent with the input values.

%%%FIG%%%%%%%%

It is clear from  Fig.~\ref{figsing} that our 
procedure works well for the simple single component systems. We notice that the 
distribution of differences between \dela output and input values is 
well fitted by a Gaussian (see Fig.~\ref{figdou}) with $\sigma = 0.23\times10^{-5}$. 
Based on central limit theorem and Gaussian distribution of the recovered values 
around the mean, we expect the accuracy of the estimated mean value to be 
$\sim\sigma/\sqrt{\rm N}$ for N measurements.
Strong blending and internal structure in the cloud will imply larger uncertainty. 

The improvement we expect from the enhanced S/N ratio in our sample 
compared to other studies is  visualized 
in Fig.~\ref{figsbyn}. Here, we show the results obtained for 
individual realizations of single component systems  
with input \dela = 0 for S/N = 30 and 70
in the top and bottom panels respectively. The shaded region in the figure
shows the mean and $\sigma$ in the distribution of recovered values.
The distribution of individual values has $\sigma$ of
0.409$\times$10$^{-5}$  and 0.239$\times$10$^{-5}$ for signal-to-noise ratios 30 and 70 respectively.
It is clear that the improvement in the S/N ratio gives 
a factor 2 improved 
accuracy in the recovered value compared to what has
been done previously. 

In all the simulations discussed till now  we have used only
strong but unsaturated Fe~{\sc ii} lines (i.e $N$(Fe~{\sc ii})$\ge 10^{13}~
{\rm cm}^{-2}$). This ensures that the absorption profiles are
well defined.
However, when the lines are weak then the flux in the line is
appreciably affected by Poisson noise. 
One can therefore infer that the 
scatter in the measurements will be larger if weak lines are used
in the analysis. We demonstrate this using simulations
of weak lines. For simplicity we have assumed input \dela =0,
 $N$(Mg~{\sc ii}) = $1.5\times10^{12}$ cm$^{-2}$ and $N$(Fe~{\sc ii}) =
5$\times10^{11}$ cm$^{-2}$. We have simulated 30 systems
by just changing the error spectrum. The results are summarized in 
Fig.~\ref{figweak}. In the left hand side panel we show the profiles
for one realization.  From this figure one can
appreciate the distortion in the absorption profile of Fe~{\sc ii}
lines caused by Poisson fluctuations. The top panel in the
right hand side shows the recovered values of \dela.
As expected the error bars are larger (one $\sigma$ of the distribution
around the mean is 2.92$\times10^{-5}$).  Typical errors
are then a factor 10 higher than that we derive with strong single
component systems. Thus if one is going to use $1/\sigma^2$
weighting of different measurements, only one single strong and
narrow component can be as important as about 100 weak systems.
When considering 30 realizations, we find a weighed mean 
of \dela = $(-0.828\pm0.508)\times 10^{-5}$ that is a 
1.6$\sigma$ deviation with respect to the zero input value.
The results shown in the bottom right panel in Fig.~\ref{figweak}
are for $N$(Fe~{\sc ii}) = 10$^{12}$ cm$^{-2}$
and $N$(Mg~{\sc ii}) = 3$\times10^{12}$ cm$^{-2}$.  The
distribution of individual measurements has 
$\sigma$~=~ $1.43\times10^{-5}$
and the weighted mean value \dela = $(0.076\pm0.273)\times 10^{-5}$. 
Thus this study clearly demonstrates that 
if the lines are weak and the profiles
are dominated by Poisson errors then a marginally significant
falls alarm detection of \dela can be obtained even when
one uses large number of systems.
This can be avoided in the analysis of real data by
carefully estimating appropriate lower limit cutoff in column density 
for the sample.  Based on our simulations and data (see below) we obtain this
cutoff {for Fe~{\sc ii} column density} to be of the order of $2\times10^{12}$ cm$^{-2}$ for
 the median S/N ratio
corresponding to our data.

\subsection{\dela measurement in strongly blended systems}

Next we consider the case of strongly blended two component systems.
Here $N$ and $b$ for the two components are chosen in the same way as
for the single component systems (see section 3.1).
The separation between the two components is taken between 3 and 5 \kms.
This means that 
the subcomponents are separated by less than the width of individual components.
The relationship between the input and the recovered \dela is shown in 
top right-hand panel of Fig.~\ref{figdou}.  This panel also shows
a typical simulated profile. There are 30 realizations
for a given input \dela. Typical error in the recovered values is
shown in the figure. It is apparent  that individual measurements 
deviate much more from the actual value compared to 
the single component case. Unlike in the case of single component
systems the deviation with
respect to the actual value can not be approximated by a single
Gaussian profile (see bottom right panel in Fig.~\ref{figdou}). 
There is an extra tail in the distribution on both sides. 
This clearly demonstrates that statistically 
significant deviations from the real value is possible for
a non-negligible fraction of the systems. 
To avoid this, we will predominantly use systems 
that are not strongly blending (i.e, for which the 
sub-component separations are larger
than the individual $b$ parameters). 

\subsection{\dela measurements from well separated blends}

\begin{figure}
%\centerline{\vbox{
\psfig{figure=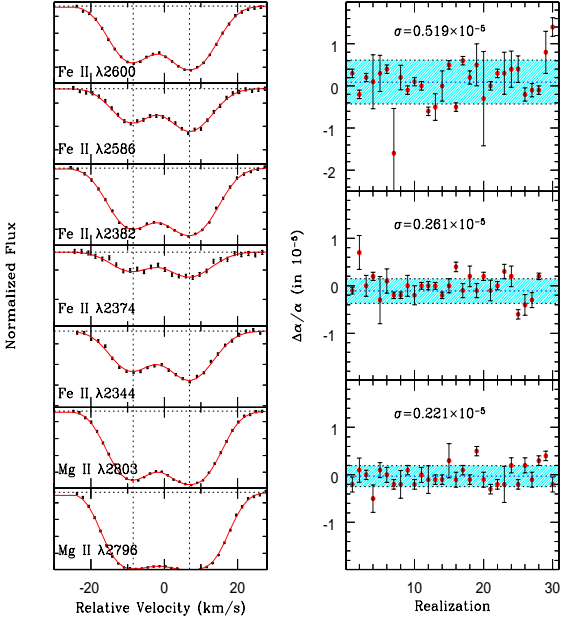,height=9.cm,width=8.cm,angle=00}
%}}
\caption{{\bf Results for well separated blends:} Left hand-side
panels show the velocity plot and the best fitted Voigt profiles
to Mg~{\sc ii} and Fe~{\sc ii} absorption lines from
one of the realizations. In all cases considered  here
the velocity
separation between the components are always more than the individual
values of the $b$ parameters. In the right hand-side panels
we compare the results 
obtained for heavily blended systems (top panel),
well separated blends (middle panel) and single component
systems (lower panel). The standard deviation measured around 
the mean in different cases is also given in the corresponding 
panels. It is clear that systems with well separated sub-components are 
as good as that of single component systems.
}
\label{figsep}
\end{figure}

The next step is to investigate the case of multicomponent systems
where subcomponents are marginally blended. That is the cases where the 
separation between individual components is larger than the
individual $b$ values. 
Results are summarized in Fig.~\ref{figsep}. The left-hand side panel
shows velocity profiles for one of the realizations. In the right 
hand-side panels we compare
results obtained for the case considered here (middle panel)
with that obtained for single component systems (lower panel) and
strongly blended systems (top panel). The $\sigma$ of
the distribution around the
mean are 0.221$\times10^{-5}$, 0.261$\times10^{-5}$ and 
0.519$\times10^{-5}$ for the single, weakly blended and 
strongly blended systems respectively. 
It is apparent that the mildly blended case gives very similar 
results as that obtained with single components.  

\subsection{Conclusions from simulations}
In short, we can conclude from these simulations that:
\begin{enumerate}
\item{} Best constraints on \dela are obtained either from  
single component systems or  well resolved multiple component systems.
\item{} Increasing the signal-to-noise ratio from S/N~=~30 to 70
increases the accuracy of \dela measurements by about a factor of two.
\item{} It is better to avoid weak lines 
while extracting \dela as their profiles can be distorted by 
Poisson noise. Thus, weak lines in the
low signal-to-noise data can result in false alarm detections of 
non-zero \dela values.
\item{} There is a non-negligible probability to derive a
statistically significant deviation from the actual value when 
one considers highly blended systems (i.e., systems where the
component separations are smaller than the individual $b$ values). 
Thus it is better to avoid complex blends in the analysis.
\end{enumerate}
In what follows we use these results as guidelines for choosing 
the systems in our sample for the final analysis.

\section{Description of the UVES sample}

\begin{table}
\caption{List of Mg~{\sc ii} \& Fe~{\sc ii} systems}
\begin{tabular}{llllll}
\hline\hline
{QSO}   &{$z_{em}$}   & {$z_{abs}$} & {Transitions}$^{p}$ & Comments \\
\hline
HE1341-1020    & 2.135  & 0.8728     & abdefghi  &  \\
               &        & 1.2778     & abcdefghi &  \\
               &        & 1.9154     & abcdefghi &  \\
Q0122-380      & 2.190  & 0.8225     & abdfghi   &  \\
               &        & 0.8585     & abefgh    &  \\
               &        & 1.2433     & abdefi    &  \\
               &        & 1.9102     & abfh      & weak  \\
%               &        &1.97233     & abfh      &  \\ 
PKS1448-232    & 2.220  & 1.5847     & abfh      & weak \\
PKS0237-23     & 2.222  & 1.1846     & abfgh     &  weak\\
               &        & 1.3650     &abcdefghi  & \\
               &        & 1.6371     & abcdefghi &  \\
               &        & 1.6575     & abcdefghi &  \\
               &        & 1.6724     & abcdefghi &  \\ 
HE0001-2340    & 2.263  & 0.4524     & abdefghi  &  \\
               &        & 0.9489     & abfh      & weak  \\
               &        & 1.5855     & abicdefgh &  \\
               &        & 1.6517     & abfi      &  weak \\
               &        & 2.1839     & abdefghi  &  \\
Q0109-3518     & 2.404  & 1.1828     & abdfghi   &  \\
               &        & 1.3499     & abcdefghi &  \\ 
HE2217-2818    & 2.414  & 0.7862     & abfh      & weak  \\
               &        & 0.9425     & abdfghi   &  \\
               &        & 1.6917     & abcdefghi &   \\
               &        & 1.6277     & abcdefghi &  weak \\
               &        & 1.5556     & abcdfghi  &    \\
Q0329-385      & 2.435  & 0.7627     & abefghi   &  \\
               &        & 1.4379     & abdhi     & weak  \\
HE1347-2457    & 2.611  & 1.4392     & abcdefghi &  \\
               &        & 1.5082     & abdfh     &  weak  \\
Q0453-423      & 2.658  & 0.7261     & abdefghi  &  \lya\\
               &        & 0.9083     & abdefghi  &  \\
               &        & 1.0394      & abh      & weak \\
               &        & 1.1492      & abdefghi &  + \\
               &        & 1.6302      & abf      & weak \\
               &        & 1.8584      &abdfh     &  \\
               &        & 2.3004      & abcdef   & +  \\
PKS0329-255    & 2.703  & 0.9926      &abdfghi   & weak  \\
Q0002-422      & 2.767  & 0.8367      &abdefghi  & \lya \\
               &        & 1.5418      &abdefghi  &  \\
               &        & 1.9888      &abdefghi  &  weak \\
               &        & 2.1678      &abcdefghi &  \\
               &        & 2.3018      &abcdefi   &  \\
HE0151-4326    & 2.789  & 0.6632      & abdefghi &  +  \\
               &        & 1.7319      & abf      &  weak \\
HE2347-4342    & 2.871  & 1.7962      & abdfgh   &  weak \\
HE0940-1050    & 3.084  & 1.0598      & abdgh    &  no anchor\\
               &        & 1.7893      & abdefghi &  +    \\
               &        & 1.9182      & adefghi &  no anchor\\
PKS2126-158    & 3.280  & 0.6631      & abdefg   &  \lya \\
               &        & 2.0225      & abdefgh  &  \\
\hline
\hline
\\
\multicolumn{5}{l}{$^{p}$Transitions as identified in Table~\ref{tabat}.}\\
\multicolumn{5}{l}{''weak'' refers to systems with 
$N$(Fe~{\sc ii})~$<$~$2\times 10^{12}$cm$^{-2}$.}\\
\multicolumn{5}{l}{''+'' indicates systems that are not used in the}\\
\multicolumn{5}{l}{~~~~~~~~analysis (see discussion).} \\
\multicolumn{5}{l}{''\lya'' refers to systems which are blended in the \lya forest.} \\
\multicolumn{5}{l}{''no anchor'' refers to systems without good anchor lines.} \\
\\
\label{tablist}
\end{tabular}
\label{tablist}
\end{table}

The Mg~{\sc ii} systems with detectable Fe~{\sc ii} absorption lines
present in our data are listed in Table.~\ref{tablist}.
The table provides the QSO name, the emission redshift
(\zem), the redshift of the absorption line systems (\zabs),
the species that are detected (notations are as in Table.~\ref{tabat}). 
No such Mg~{\sc ii} 
systems are detected  along 2 sight lines 
(toward HE~$1158-1843$ at \zem =2.449
 and Q~$0420-388$ at \zem = 3.177).
Last column of the table is meant for 
comments on the systems.

In total, there are 50 systems with detected Mg~{\sc ii} and 
Fe~{\sc ii} lines 
at a detection limit of $\sim$$1.5\times 10^{11}$ cm$^{-2}$
for the strongest Fe~{\sc ii}$\lambda$2382 line.
The corresponding limit for Fe~{\sc ii}$\lambda$2374 is
1.5$\times 10^{12}$ cm$^{-2}$. 
In the following we discriminate systems with 
$N$(Fe~{\sc ii})~$<2\times10^{12}$ cm$^{-2}$ 
as simulations have shown that  
spurious effects can affect results in case the lines are
weak.
Therefore systems with $N$(Fe~{\sc ii}) less than this
limit (15 systems) are marked as ``weak'' in the table and are
not  considered in our main analysis. However, for completeness,
we present the results obtained from these systems 
as well in the concluding section.

Two systems at \zabs = 1.0598 and \zabs~=~1.9183 
toward HE~0940$+$1050 could
not be considered in our analysis because there are no good anchor 
lines in these systems. In the former system only 
Mg~{\sc ii}$\lambda$2796 is present in the spectrum. 
In the case of  the \zabs=1.9183 system, 
Si~{\sc ii} and Al~{\sc ii} are blended with \lya lines and
Mg~{\sc ii} is heavily affected by atmospheric lines. 
In addition, at \zabs = 0.726176 toward Q~0453$-$423, \zabs = 0.8367 toward
Q0002-422 and \zabs = 0.6631 toward PKS 2126$-$158, 
the Fe~{\sc ii} lines are heavily blended with \lya systems
and are therefore marked as ``Ly$\alpha$'.
Thus we are left with 29 systems that have strong Fe~{\sc ii}
lines, good anchor lines and are not contaminated by  
\lya or atmospheric absorption. These systems form the basic dataset 
that we use for measurement of \dela. 

In summary, based on the results from the simulations, we apply the following 
selection criteria to derive reliable \dela:
\begin{enumerate}
\item{} We consider only lines with  similar ionization potentials 
(Mg~{\sc ii}, Fe~{\sc ii}, Si~{\sc ii} and Al~{\sc ii}) as they are 
most likely to originate from similar regions in the cloud.
\item{} We avoid absorption lines that are contaminated by atmospheric lines.
\item{} We consider only systems that have 
$N$(Fe~{\sc ii})$\ge 2\times10^{12}$ cm$^{-2}$ which ensures
that all the standard Fe~{\sc ii} multiplets are detected 
at more than  5$\sigma$ level.
\item{} We demand that at least one of the anchor lines is
not saturated so that the redshift measurement is robust.
\item{} We also avoided sub-DLAs (i.e $N$(H~{\sc i})$\ge10^{19}$ cm$^{-2}$) 
as these systems may have ionization and chemical
inhomogeneities.
\item{} We do not consider strongly saturated systems with large velocity spread
(complex blends); however in such systems
whenever we find a well detached satellite components we 
include these components in the analysis.
\item{} finally, based on the component structure resulting from
the Voigt profile fits of systems that are not complex
blends, we retain only 
systems for which the majority of components are separated
from its neighboring components by more than the $b$ parameters. 
\end{enumerate}
Application of the above conditions resulted in 23 systems
on which to perform the measurement (six single component systems, 
six well separated doubles, six systems with three components 
with at least one well detached from the rest, and  five 
systems with more than 3 components) .
\par\bigskip\noindent
\section{Notes on individual systems in our sample}
\subsection{Systems along the line of sight toward HE~1341-1020}
There are 3 Mg~{\sc ii} systems with Fe~{\sc ii} along this
sight line.  All these three systems are strong enough to be
included in our sample.
\subsubsection{\zabs = 0.8728 system toward HE~1341-1020}
\begin{figure}[h]
\centerline{\vbox{
\psfig{figure= 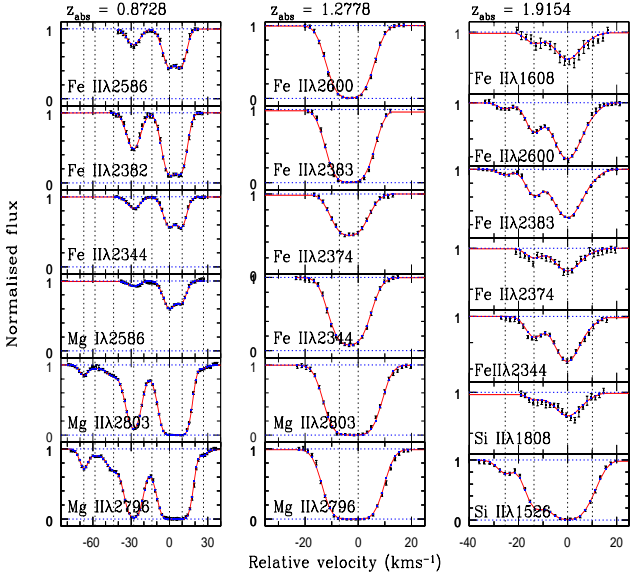,height=8.cm,width=9.cm,angle=270.}
}}
\caption[]{Voigt profile fits to the absorption systems  
seen along the line of sight toward \zem = 2.135 QSO HE~1341-1020.
The vertical dotted lines mark the locations of subcomponents.
}
\label{velp1}
\end{figure}
The Voigt profile fits to some of the lines from this system are
shown in Fig.~\ref{velp1}. This system shows  Al~{\sc ii}, Mg~{\sc i}
$\lambda$2852, Ca~{\sc ii}$\lambda\lambda$3934,3969 lines in addition 
to Mg~{\sc ii} and Fe~{\sc ii} lines. The Mg~{\sc ii} profile is
fitted with 8 components, the three stronger of which 
show absorption in other species. Fe~{\sc ii}$\lambda$2600 line is 
blended with absorption lines from other intervening systems and 
the Fe~{\sc ii}$\lambda$2374 profile is affected by a bad pixel.  
In order to avoid unwanted systematics we do not consider these lines 
for the $\alpha$ measurement analysis. However we use these profiles 
to check the consistency in the number of components and the level of 
saturation in each component. The Al~{\sc ii} line is affected by
noise and Ca~{\sc ii} lines are too weak to give any 
meaningful constraints. We therefore use 
Fe~{\sc ii}$\lambda\lambda\lambda$2344,2382,2386, 
Mg~{\sc ii}$\lambda\lambda$2796,2803 and Mg~{\sc i}$\lambda$2586.
\subsubsection{\zabs = 1.2778 system toward HE~1341-1020}
The absorption profile of thi-s system is spread over 200 \kms.
Absorption lines of Mg~{\sc i}, Mg~{\sc ii}, 
Al~{\sc ii}, Al~{\sc iii},
Fe~{\sc ii}, Ca~{\sc ii}, Si~{\sc ii}, Mn~{\sc ii}, Ni~{\sc ii}, and
Ca~{\sc ii} are detected. The core is strongly blended over
100~\kms but a well detached narrow satellite is seen $\sim$100~km~s$^{-1}$ 
away. We have used Fe~{\sc ii} and Mg~{\sc ii} absorptions from this sharp 
component for the measurement of \dela. We do not use Mn~{\sc ii}, Ni~{\sc ii}
and Ca~{\sc ii} because they are weak. Si~{\sc ii} and Al~{\sc ii}
lines fall in the \lya  forest and the profile of
Fe~{\sc ii}$\lambda$2586 is affected by the wing of a contaminating 
neighboring line. The satellite component is well
fitted with two components (see middle panel in Fig.~\ref{velp1}).
Even though the two components are merged, we notice that the 
components are separated by more than the velocity dispersion of
the individual component. 
\subsubsection{\zabs = 1.9154 system toward HE~1341-1020}
This system shows absorption lines from Mg~{\sc ii}, Mg~{\sc i},
Al~{\sc ii}, Al~{\sc iii}, C~{\sc iv}, Si~{\sc ii}, and Si~{\sc iv}. 
High-ionization lines are weak and their profiles  are very
different from that of low-ionization lines. Mg~{\sc ii}
lines are saturated and blended with atmospheric absorption lines.
Thus, we use Si~{\sc ii}$\lambda$1526, Si~{\sc ii}$\lambda$1808
and Fe~{\sc ii} lines for \dela measurement. 
We do not use Fe~{\sc ii}$\lambda$2586 however as its profile
is affected by noise. The system is
fitted with 4 distinct components that are clearly visible in
the profiles of the weak lines (see right panel in
Fig.~\ref{velp1}). 
Even though the overall fit is good, one can notice 
that the Si~{\sc ii}$\lambda$1808 line profile is apparently not 
reproduced well. This is due to the effect of noise in a single pixel 
in the red wing of the strongest absorption component.
This shows how delicate the measurement can be.
\subsection{Systems along the line of sight toward Q~0122$-$380}
\begin{figure*}
\centerline{\vbox{
\psfig{figure= 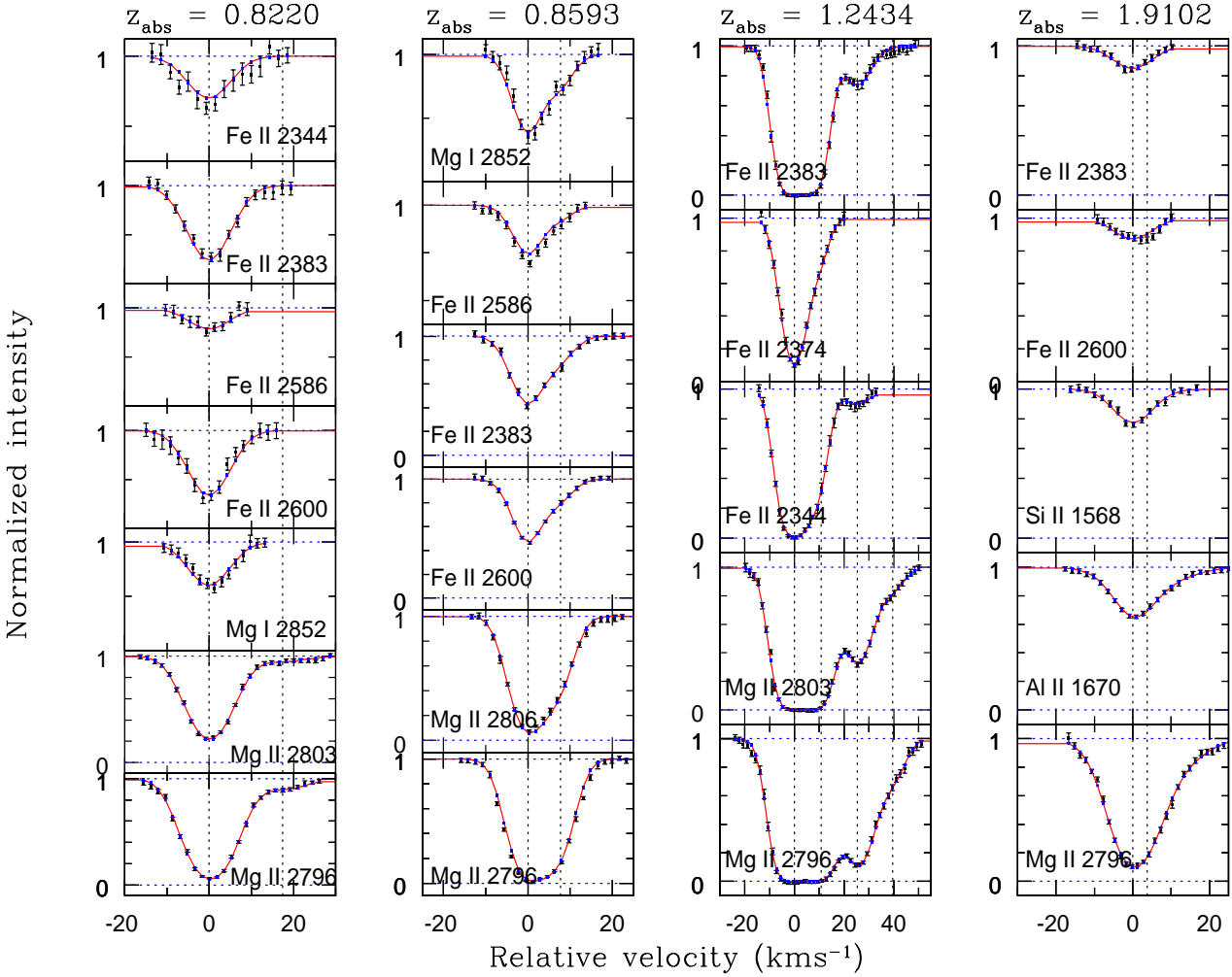,height=14.cm,width=18.cm,angle=270.}
%comph1.ps
}}
\caption[]{Voigt profile fits to the four absorption systems 
seen along the line of sight toward \zem = 2.190 QSO Q~0122$-$380.
The vertical dotted lines mark the locations of subcomponents.
}
\label{velp2}
\end{figure*}
There are 4 Mg~{\sc ii} systems along this line of sight.
The system at \zabs = 1.9102 has weak Fe~{\sc ii} absorption
lines and do not fulfill our criteria to enter our main sample.
\subsubsection{\zabs = 0.8221 system toward Q~0122$-$380}
This system is defined by narrow Mg~{\sc ii}, Mg~{\sc i} and
Fe~{\sc ii} absorption lines. All the lines are well fitted by a 
single component. An extra component in the red wing is required for
the Mg~{\sc ii} doublets. The Voigt profile fits to the system
are shown in Fig.~\ref{velp2}. 
\subsubsection{\zabs = 0.8593 system toward Q~0122$-$380}
This system is defined by Mg~{\sc ii}, Mg~{\sc i}, Al~{\sc ii}, Al~{\sc iii},
Ca~{\sc ii}, and Fe~{\sc ii} absorption lines. 
As the spectral range in which Al~{\sc ii} absorption is redshifted
is of lower signal to noise and Ca~{\sc ii} doublets and 
Fe~{\sc ii}$\lambda2474$ are weak we have not used these lines for the \dela analysis. 
Resulting Voigt profile fits are shown in Fig.~\ref{velp2}. We notice that
the Fe~{\sc ii}$\lambda$2586 line is slightly under-predicted in our best model fit. 
However other three Fe~{\sc ii} lines are very well fitted.
We also notice that $\chi^2$ per degree of freedom 
is very good when using different $b$ parameters for Fe~{\sc ii} 
and Mg~{\sc ii}. The result on \dela presented for this system is therefore
obtained with different $b$ parameters for Mg~{\sc ii} and Fe~{\sc ii} 
lines. 
\subsubsection{\zabs = 1.2433 system toward Q~0122$-$380}
Mg~{\sc ii} absorption profiles in this system is spread over
$\sim70$ \kms. The components are heavily blended. The strongest
Mg~{\sc ii} component shows Mg~{\sc i}, Al~{\sc iii}, Si~{\sc ii},
Si~{\sc iv}, C~{\sc iv}, Ca~{\sc ii}, Ni~{\sc ii} and Fe~{\sc ii}. 
Fe~{\sc ii}$\lambda$2586 and Fe~{\sc ii}$\lambda$2600 lines are not
covered by our spectrum. Three Fe~{\sc ii} lines and the Mg~{\sc ii} 
doublet can be fitted with 4 components. Ni~{\sc ii}$\lambda$1704
line is blended and  Ni~{\sc ii}$\lambda$1741 is weak and falls in
the noizy region of the spectrum. We notice small profile inconsistencies
between the Ca~{\sc ii} doublets. Thus we avoid using Ni~{\sc ii}
and Ca~{\sc ii} lines in the analysis.
The results are summarized in Fig.~\ref{velp2}. 
\subsection{Systems along the line of sight toward PKS 0237$-$23}
\begin{figure*}
\centerline{\vbox{
\psfig{figure= 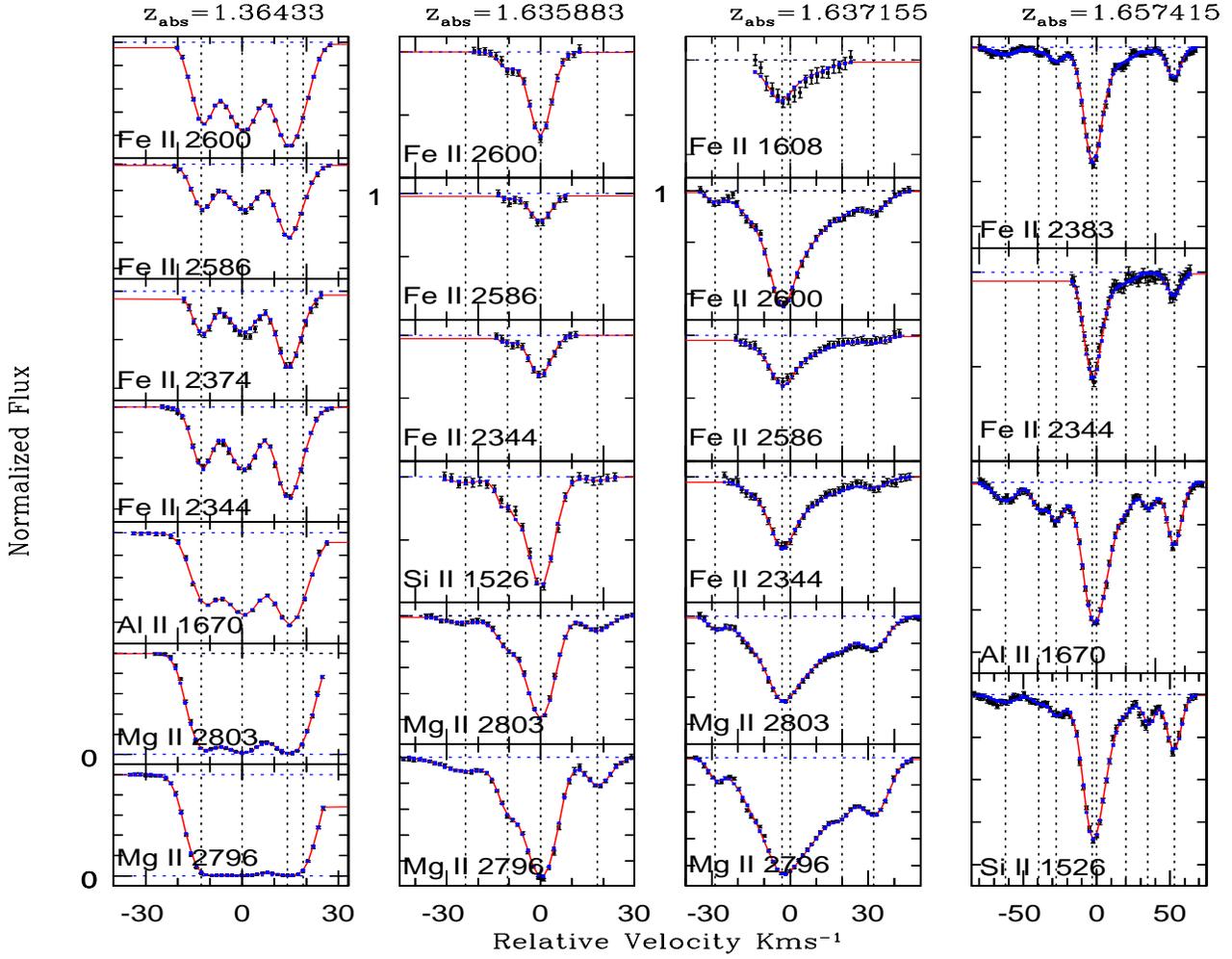,height=14.cm,width=18.cm,angle=0.}
%comph1.ps
}}
\caption[]{Voigt profile fits to the five absorption systems 
seen along the line of sight toward \zem = 2.222 PKS 0237$-$23.
The vertical dotted lines mark the locations of subcomponents.
}
\label{velp0237}
\end{figure*}
This QSO is known for the presence of a super cluster of C~{\sc iv}
absorption lines at \zabs $\sim$ 1.6. Our spectrum reveals 5 Mg~{\sc ii}
systems one of which at \zabs = 1.1846 has very weak Fe~{\sc ii} line 
and is not considered for our main analysis.
\subsubsection{\zabs = 1.3650 toward PKS 0237$-$23}
This system shows absorption lines of Fe~{\sc ii}, Al~{\sc ii},
Al~{\sc iii}, C~{\sc i}, C~{\sc i}$^*$, C~{\sc ii}, C~{\sc ii}$^*$,
Si~{\sc ii}, Si~{\sc iv}, Ca~{\sc ii}, Mg~{\sc i} and Mg~{\sc ii}
lines.  The \lya line is just below our wavelength coverage.
The absorption profiles of strong resonance lines spread 
over 230 \kms.  The main component has large number of 
sub-components heavily blended. There is a well detached
satellite component that could be used for the \dela
measurements.
Presence of C~{\sc i} suggests that this could possibly be
a sub-DLA. In such systems inhomogeneities are generic. Thus
it is better to avoid them in the \dela measurements. 
We notice that this system happens to be in common with 
our data and that of Murphy et al.(2003). Thus we perform
the \dela measurement in the satellite components.
We use Al~{\sc ii}, all Fe~{\sc ii} lines 
(apart from Fe~{\sc ii}$\lambda$2383), and
Mg~{\sc ii} lines for our analysis.
Murphy et al (2003) has found \dela = -0.19$\pm0.5$ for
this system. Unlike our case Murphy et al have used
the whole profile of Al~{\sc ii}, Si~{\sc ii}, Al~{\sc iii},
and two Fe~{\sc ii} lines. Our analysis gives the best fitted 
value of 0.0$\pm0.1$.
This is consistent with that measured by
Murphy et al and our final result. The increased accuracy in
our measurement is contributed by the enhanced S/N and our
choice of the well detached and unblended satellite component
for the analysis.
For the reason
mentioned above we have avoided this system in our analysis.
\subsubsection{\zabs = 1.6371 toward PKS 0237$-$23}
The absorption profiles of this system is spread over $\sim$280 \kms.
Absorption lines from Mg~{\sc i}, Mg~{\sc ii}, Al~{\sc ii}, Al~{\sc iii}
Si~{\sc ii} and Fe~{\sc ii} are detected.  Even though C~{\sc iv}
and Si~{\sc iv} absorption lines are detected the system is dominated
by low ionization species. There are two distinct velocity components
at \zabs = 1.63588 and 1.63715 that we use for \dela measurements.
The Voigt profile fits to these systems are shown in Fig.~\ref{velp0237}.
\subsubsection{\zabs = 1.6574 toward PKS 0237$-$23}
This system shows absorption lines from Mg~{\sc ii}, Mg~{\sc i}, Al~{\sc ii},
Al~{\sc iii}, Si~{\sc ii}, C~{\sc iv}, Si~{\sc iv} in addition to
Fe~{\sc ii}. The low  and high-ionization lines show distinct profiles.
C~{\sc iv} and Si~{\sc iv} absorption lines are spread over 400 km/s. We use
the uncontaminated Fe~{\sc ii} lines with Si~{\sc ii} and Al~{\sc ii}
lines as anchors. Mg~{\sc ii} line is heavily blended and has large
number of extra components compared to Fe~{\sc ii} and Mg~{\sc i} is
weak. The results of the Voigt profile fits are shown in 
Fig.~\ref{velp0237}. 
\subsubsection{\zabs = 1.6724 toward PKS 0237$-$23}
This is one of the few sub-DLAs (damped \lya systems)
that are detected in the large programme sample. 
Mg~{\sc i}, Mg~{\sc ii}, C~{\sc i},C~{\sc i}$^*$,
C~{\sc ii},C~{\sc ii}$^*$, Si~{\sc ii}, Si~{\sc iv}, C~{\sc iv}, Ni~{\sc ii},
Zn~{\sc ii}, Cr~{\sc ii}, Mn~{\sc ii} and Fe~{\sc ii} lines are
present. Most of the standard resonance lines are saturated. 
Zn~{\sc ii}, Mn~{\sc ii}, Cr~{\sc ii} and Ni~{\sc ii} absorption lines are 
weak and therefore will not give good constraints. The presence of the whole
range of ionization states and of fine-structure absorption lines strongly
suggests strong inhomogeneities in the gas. To keep our analysis unbiased
we did not use this sub-DLA system in our analysis.                
\subsection{Systems along the line of sight toward HE~0001-2300}
There are 5 Mg~{\sc ii} systems detected along this sight line.
Two of them at \zabs = 0.9489 and 1.6517 are weak systems and are not
considered in our main sample.
\begin{figure*}
\centerline{\vbox{
\psfig{figure= 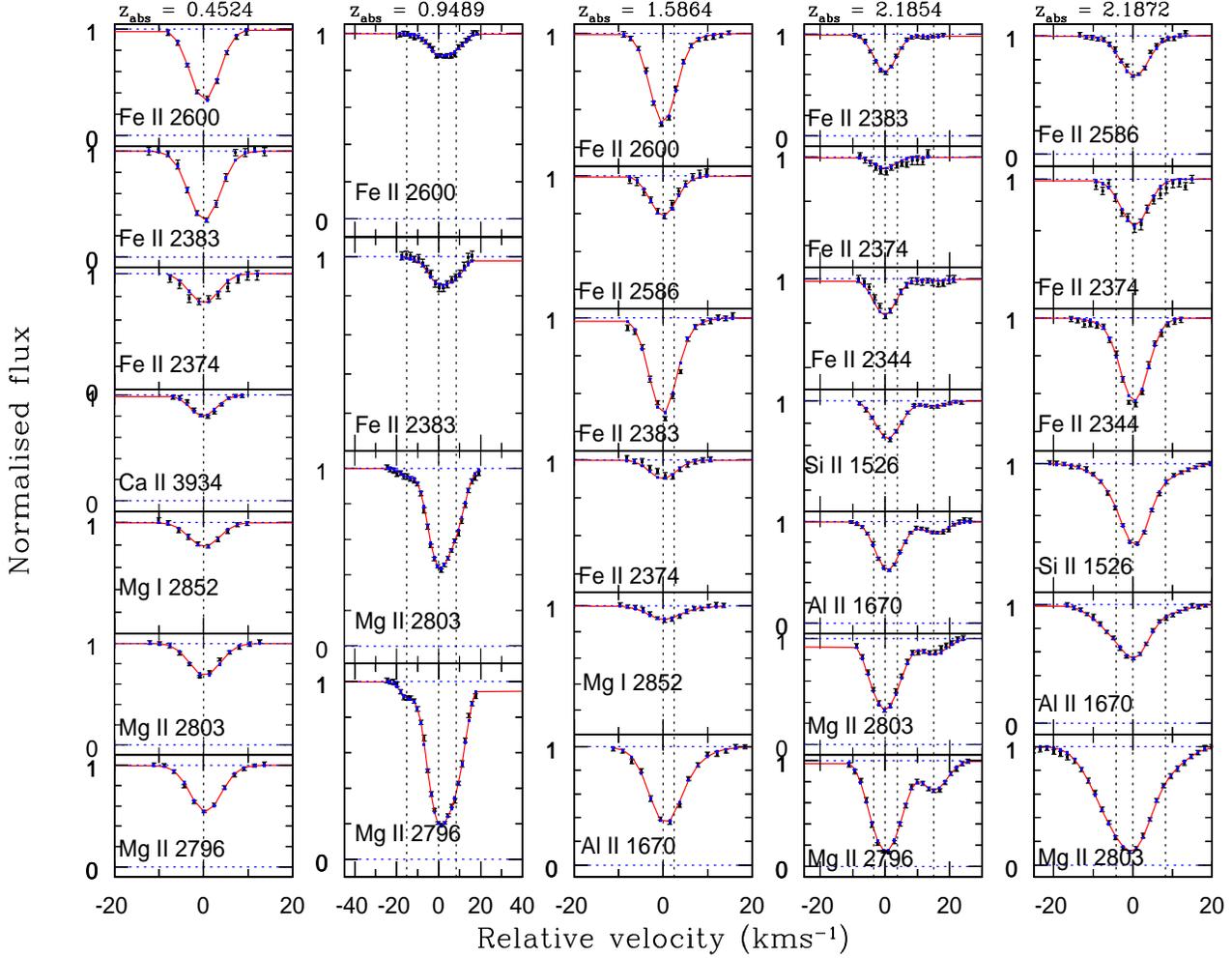,height=14.cm,width=18.cm,angle=270.}
%comph1.ps
}}
\caption[]{Voigt profile fits to the five absorption systems 
seen along the line of sight toward \zem = 2.263 QSO HE~0001$-$2300.
The vertical dotted lines mark the locations of subcomponents.
}
\label{velp0001}
\end{figure*}
\subsubsection{\zabs = 0.4524 system toward HE~0001$-$2300}
This system has two very narrow components separated by $\sim70$ \kms.
The low redshift component shows Mg~{\sc ii}, Mg~{\sc i}, Fe~{\sc ii}
and Ca~{\sc ii}$\lambda$3934 absorption lines.  
The high redshift component has very weak Fe~{\sc ii} lines. 
Thus we use only the low redshift component for \dela measurement. 
Voigt profile fits to the absorption lines from this component are shown 
in Fig.~\ref{velp0001}. Fe~{\sc ii}$\lambda2344$ and Fe~{\sc ii}$\lambda2586$ 
are blended with other absorption lines and were not used in the analysis.
\subsubsection{\zabs = 1.5855 system toward HE~0001$-$2300}
Like the \zabs = 0.4524 system  this system has two well detached 
components separated by $\sim110$ \kms. Mg~{\sc ii}, Mg~{\sc i}, Fe~{\sc ii}
Al~{\sc ii}, Al~{\sc iii}, C~{\sc iv}, Si~{\sc ii} and Si~{\sc iv} absorption
lines are seen in both components. However Fe~{\sc ii} lines are strong
only in the high-$z$ component.  We notice that Mg~{\sc ii}$\lambda2803$ is
contaminated by an atmospheric line. As Si~{\sc ii}$\lambda$1526 is also blended,
we use Al~{\sc ii} and Mg~{\sc i} absorption lines as anchors.
Even though the overall fit to the data is good, our profiles 
consistently over predict the blue wing of Fe~{\sc ii}$\lambda$2374.
This is probably due to one single pixel in the wing being affected
by noise. As the two components that are required to fit the profiles are
very close to each other this system is not considered for the
final analysis.
\subsubsection{\zabs = 2.1839 system toward HE~0001$-$2300}
Absorption profiles produced by this system are spread over 400 \kms.
Mg~{\sc i}, Mg~{\sc ii}, Al~{\sc ii}, Al~{\sc iii}, Si~{\sc ii},
Si~{\sc iv} and C~{\sc iv} absorption lines are detected. Fe~{\sc ii}
lines are strong only in two of these components (at \zabs = 2.1854 and
\zabs = 2.1872). As Fe~{\sc ii}$\lambda2383$, Fe~{\sc ii}$\lambda2586$
and Fe~{\sc ii}$\lambda2600$ absorption lines fall in the wavelength
range affected by atmospheric absorption, we consider only
the clean lines in each case. Results of profile fitting are 
shown in Fig.~\ref{velp0001}. 
\subsection{ Systems along the line of sight toward Q0109$-$3518}
There are two Mg~{\sc ii} system seen along the line of sight toward
this QSO. We used both of them for \dela measurement.
\begin{figure*}
\centerline{\vbox{
\psfig{figure=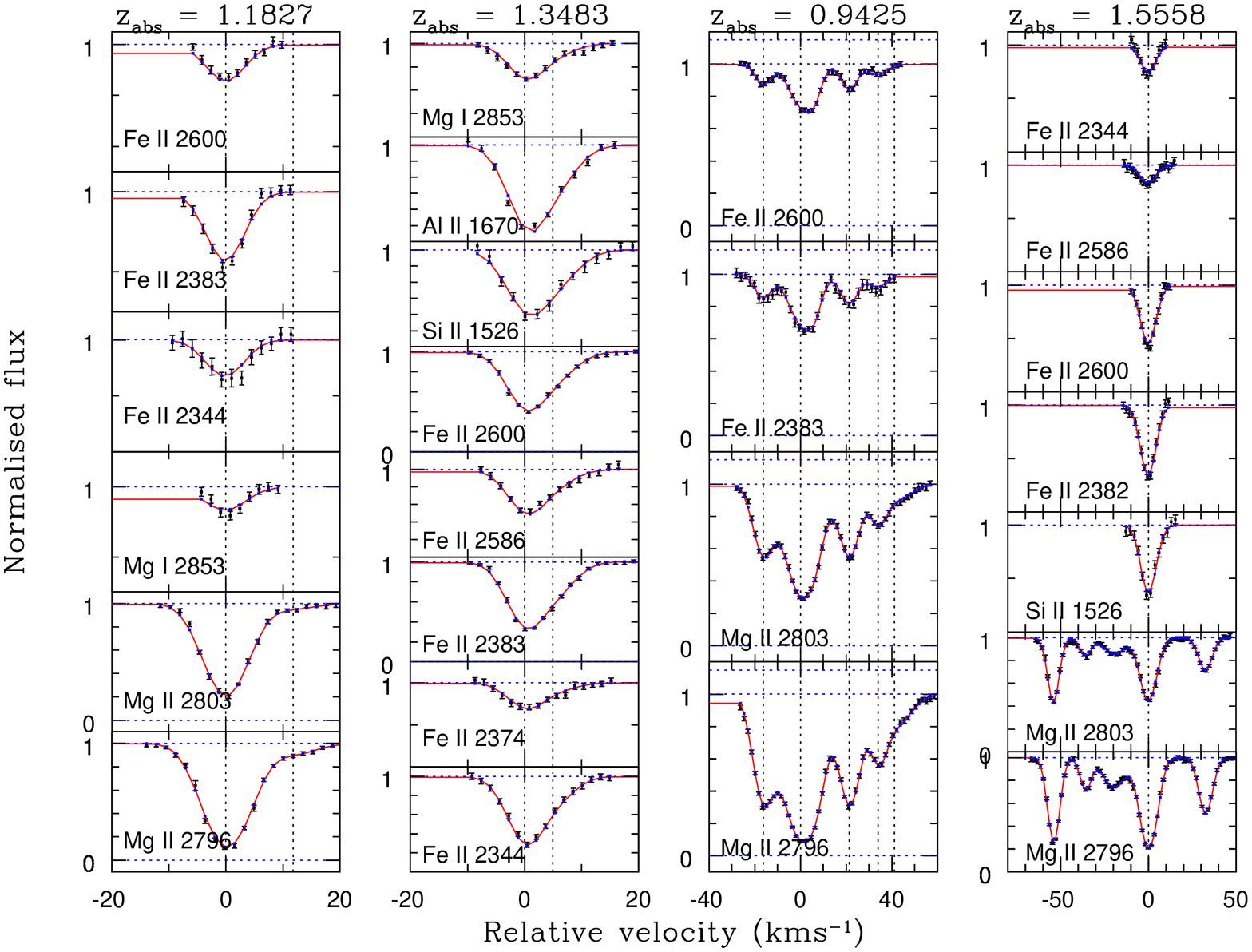,height=15.cm,width=18.cm,angle=270.}
%comph1.ps
}}
\caption[]{Voigt profile fits to two of the absorption  systems
seen along the line of sight toward \zem = 2.404  Q0109$-$3518
(first and second column from the left) and \zem = 2.414 
HE $2217-2818$ (columns 3 and 4 from the left).
The vertical dotted lines mark the locations of subcomponents.
}
\label{velp0109}
\end{figure*}
\subsubsection{\zabs = 1.1827 system toward Q0109$-$3518}
This system shows absorption lines from Mg~{\sc ii}, Mg~{\sc i}
and Fe~{\sc ii}. The Fe~{\sc ii} and Mg~{\sc i} lines
are fitted with a single component but Mg~{\sc ii} requires
additional components in the red wing. Results of Voigt
profile fitting are shown in Fig.~\ref{velp0109}. 
\subsubsection{\zabs  = 1.3483 system toward  Q0109$-$3518}
This is a strong Mg~{\sc ii} systems with heavily blended
absorption profiles spread over 300 \kms. 
Absorption lines from Mg~{\sc i}, Mg~{\sc ii}, Fe~{\sc ii}, Al~{\sc ii},
Al~{\sc iii}, Si~{\sc ii}, Si~{\sc iv} and C~{\sc iv} are detected.
There are two well detached satellite components. We used
the \zabs = 1.3483 satellite for \dela measurement.  As Mg~{\sc ii}
is blended, Si~{\sc ii}, Al~{\sc ii} and Mg~{\sc i} lines are used as
anchors. For the other satellite the anchor lines are all
blended.
% in the wings.
Results of Voigt profile fitting are presented in Fig.~\ref{velp0109}.
\subsection {Systems along the line of sight toward \zem = 2.414
HE~2217-2818}
There are five Mg~{\sc ii} systems along the line of sight toward
this QSO. Three of these systems (at \zabs = 0.7862, 1.6277 and 
1.5556) are weak and are not considered in the main sample. The other two
systems were used for \dela measurement.
\subsubsection {\zabs = 0.9425 system toward HE~2217-2818}
This system shows absorption lines from Mg~{\sc ii} and Fe~{\sc ii}. 
There are two distinct subsystems 
that are separated by $\sim 200$ \kms. Only the high redshift subsystem
shows Fe~{\sc ii} absorption. Fe~{\sc ii} lines are fitted
with four well separated components and Mg~{\sc ii} requires an
extra component in the red wing. 
We notice that the goodness of the fit in this system is decided
by how well we model the Mg~{\sc ii} profiles.
Results of profile fitting are summarized in Fig.~\ref{velp0109}.
\subsubsection {\zabs = 1.5556 system toward HE~2217-2818}
This system shows absorption due to Mg~{\sc ii}, Al~{\sc ii},
Si~{\sc ii}, Si~{\sc iv}, C~{\sc iv} and Fe~{\sc ii}. 
The Mg~{\sc ii} absorption profiles are spread over $\sim$ 100\kms.
However only the central velocity component shows other
singly ionized species (see Fig.~\ref{velp0109}). 
Surprisingly this component also shows absorption due to
O~{\sc i}. High ionization lines such as C~{\sc iv}, Si~{\sc iv}
are weak and Al~{\sc iii} is absent.
\subsection{ Systems along the line of sight toward 
\zem = 2.435 QSO Q0329$-385$}
\begin{figure*}
\centerline{\vbox{
\psfig{figure=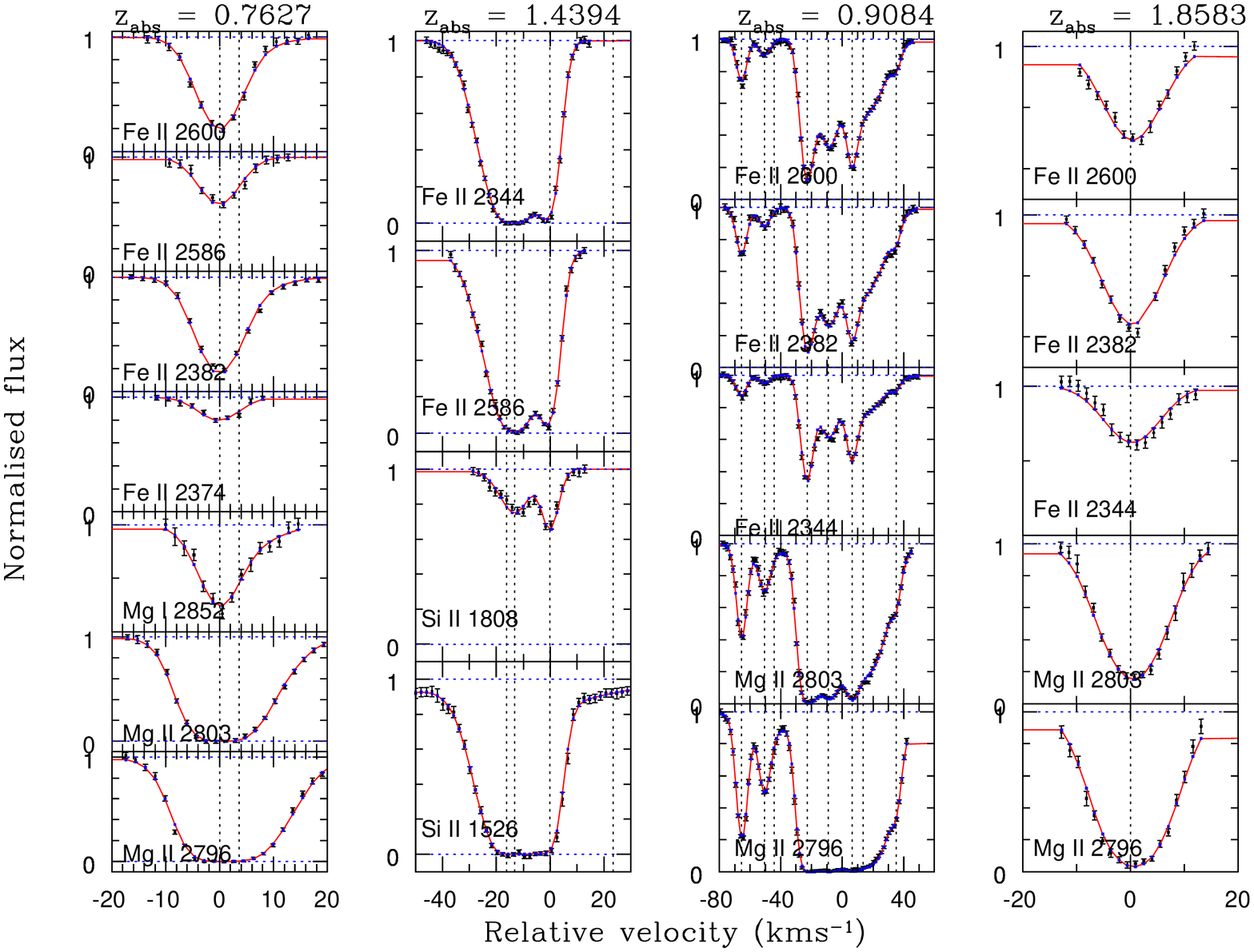,height=15.cm,width=18.cm,angle=270.}
%comph1.ps
}}
\caption[]{Voigt profile fits to the absorption systems
seen along the line of sight toward \zem = 2.435  QSO Q0329$-$385
(first column from the left), \zem = 2.611 QSO
HE $1347-2457$ (column 2 from left), and \zem = 2.658
Q $0453-423$  (columns 3 and 4 from the left) .
The vertical dotted lines mark the locations of subcomponents.
}
\label{velp0329}
\end{figure*}
There are two absorption systems detected along this line of sight. 
The \zabs = 1.43799 system is weak and is not considered in our sample. 
\subsubsection{\zabs = 0.7627 system toward Q0329$-385$}
This system shows absorption lines due to Mg~{\sc i}, Mg~{\sc ii},
and Fe~{\sc ii}.  The Mg~{\sc ii} absorption profile is spread over
200 \kms. However the other lines are detected only in the
central component.  The Voigt profile fits to this component
is shown in Fig.~\ref{velp0329}.  The profiles are fitted with
two components and as the velocity separation is smaller than 
the largest $b$ value of the components we do not consider this
system in our study.
\subsection{ Systems along the line of sight toward Q~0453$-423$}
There are five systems with detected  Fe~{\sc ii} and Mg~{\sc ii} absorption lines.
As most of the Fe~{\sc ii} absorption lines of the  \zabs = 0.72617 
system are redshifted in the wavelength range contaminated by intergalactic 
Ly-$\alpha$ H~{\sc i} absorptions we do not use this system for \dela measurement.
The absorption profiles of the \zabs = 1.1492 system is spread over 400\kms. 
The Mg~{\sc ii} and Fe~{\sc ii} lines for which we know the $q$ values are
all saturated.  The profiles of Mg~{\sc i}, Ca~{\sc ii} and Mn~{\sc ii}
suggest a very complex blend. Thus we do not consider this system
for \dela measurement. Finally, the system at \zabs = 1.0394 show only 
Fe~{\sc ii}$\lambda$2600 absorption and is therefore not considered in the analysis. 
\subsection{\zabs = 0.9083 system along the line of sight toward Q~0453$-$423}
This system shows detectable absorption from Mg~{\sc i}, Mg~{\sc ii}, 
Fe~{\sc ii}, Ca~{\sc ii} and Al~{\sc ii}. 
Fe~{\sc ii}$\lambda$2586 and Fe~{\sc ii}$\lambda$2374 are contaminated 
by absorptions from other systems. Based on the Fe~{\sc ii} profiles 
the system is decomposed in to eight subcomponents. 
The profile fits to the Mg~{\sc ii} and Fe~{\sc ii}
lines free of contamination are shown in Fig.~\ref{velp0329}.
\subsection{\zabs = 1.8583 system along the line of sight toward Q~0453$-$423}
This is a narrow single component absorption system detected in 
Mg~{\sc ii}, Al~{\sc ii}, Al~{\sc iii}, Si~{\sc ii}, Si~{\sc iv} and 
Fe~{\sc ii}. Results of the single component 
Voigt profile fit to three prominent Fe~{\sc ii} lines and the 
Mg~{\sc ii} doublet are shown in Fig.~\ref{velp0329}.
\subsection{\zabs = 2.3004 system along the line of sight toward Q~0453$-$423}
This system shows absorption lines of Mg~{\sc ii}, Fe~{\sc ii} 
and Al~{\sc ii}. Fe~{\sc ii}$\lambda$2586 and Fe~{\sc ii}$\lambda$2600
are not covered by our spectrum. Fe~{\sc ii}$\lambda$2374 is very weak.
The signal-to-noise ratio in the spectral range where other 
Fe~{\sc ii} lines are seen is $\sim 30$ only. 
In addition, both Mg~{\sc ii} lines are
contaminated by atmospheric absorption. There is also profile
inconsistencies between the weak Al~{\sc ii} lines (a potential
anchor) and Fe~{\sc ii} lines. Thus we do not consider this system
in our study.
\subsection{Absorption systems along the line of sight toward 
\zem = 2.767 PKS 0002$-$422}
\begin{figure*}
\centerline{\vbox{
\psfig{figure= 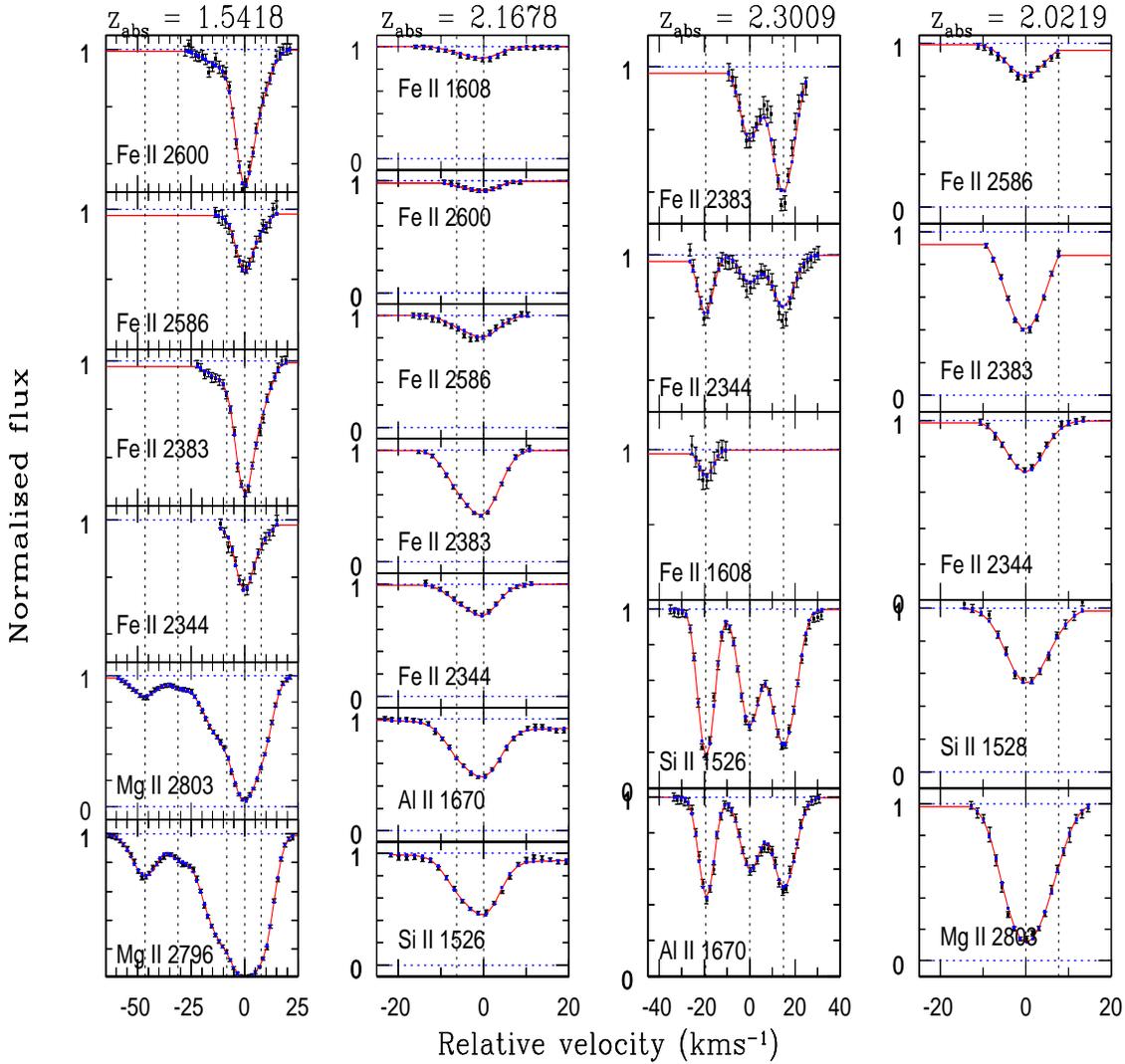,height=15.cm,width=16.cm,angle=270.}
%comph1.ps
}}
\caption[]{Voigt profile fits to two of the absorption  systems
seen along the line of sight toward \zem = 2.767  Q~0002$-422$
(First three columns from the left) and \zem = 3.280 QSO
PKS~2126$-158$ (last column).
Vertical dotted lines mark the locations of subcomponents.
}
\label{velp0002}
\end{figure*}
There are 5 Mg~{\sc ii} systems with detected Fe~{\sc ii} absorption lines
along this line of sight.
The \zabs = 0.8367 system is a very strong Mg~{\sc ii} system whose absorption
profile is spread over $\sim$600 \kms. As the corresponding absorption profiles are 
heavily blended we do not consider this system in our analysis as is the case for
the \zabs = 1.9885 system in which Fe~{\sc ii} lines are too weak. The remaining
3 systems are considered in our main analysis.
\subsubsection{\zabs = 1.5418 system toward Q~0002$-422$}
This system is defined by Mg~{\sc ii}, Mg~{\sc i}, Al~{\sc ii}, Al~{\sc iii},
and Fe~{\sc ii} absorption lines.  The Mg~{\sc ii} absorption lines are
spread over $\sim$180 \kms.  The other low ionization lines are seen only
in the strongest central component. Voigt profile fits to the 
absorption lines are shown in Fig.~\ref{velp0002}.
\subsubsection{\zabs = 2.1678 system toward Q~0002$-422$}
This system is defined by Mg~{\sc ii}, Mg~{\sc i}, Al~{\sc ii}, Al~{\sc iii},
Fe~{\sc ii}, Si~{\sc ii}, Si~{\sc iv} and C~{\sc iv} absorption lines.
Like most systems in our sample the high and low ionization profiles
have different profiles. the Mg~{\sc ii} absorption profile is spread over 
80~\kms. However only the strongest central component shows other
low ionization absorption lines. As the central Mg~{\sc ii} component
is blended in the red wing with other components we use Si~{\sc ii}
and Al~{\sc ii} as anchors.  The presence of Fe~{\sc ii}$\lambda$1608
with negative $q$ coefficient makes this system very sensitive to 
\dela. The system is fitted with two components and the velocity
separation between  them is larger then individual $b$ values.

\subsubsection{\zabs = 2.3008 system toward Q~0002$-422$}
This system is defined by Mg~{\sc ii}, Mg~{\sc i}, Al~{\sc ii}, Al~{\sc iii},
Fe~{\sc ii}, Si~{\sc ii}, Si~{\sc iv} and C~{\sc iv} absorption lines.
Like most systems in our sample the high and low ionization profiles
have different profiles. the Mg~{\sc ii} absorption profile is spread over 
200~\kms. The main component is complex and blended. There is a well
detached satellite in the blue wing of the profile. We use this
for our \dela measurement. We use the unblended Fe~{\sc ii} lines
and Si~{\sc ii}$\lambda1526$ and Al~{\sc ii} lines for our analysis.
The three components that fit the profile are well resolved.

\subsection{Absorption systems along the line of sight toward 
\zem = 3.280  QSO PKS~2126$-$158}
Two Mg~{\sc ii} systems are detected along the line of sight 
toward this QSO. The \zabs = 0.6631 system shows absorption lines from
Mg~{\sc ii}, Mg~{\sc i}, Fe~{\sc ii}, and Ca~{\sc ii}. All the lines
are heavily blended with intervening \lya absorption lines owing
to the large redshift of the QSO. We do not consider this system
for our analysis.
\subsubsection{\zabs = 2.0225 system toward  PKS~2126$-$158}
This system shows absorption due to Mg~{\sc ii}, Si~{\sc ii}
and Fe~{\sc ii}. The Mg~{\sc ii} profile is spread over 140 \kms. 
The strongest Mg~{\sc ii} component is saturated. We use the
narrow satellite component in the low redshift side,
at \zabs = 2.02192  for \dela measurement. As the Mg~{\sc ii}$\lambda$2796
is affected in the blue wing, we use Si~{\sc ii} and 
Mg~{\sc ii}$\lambda$2803 as anchors. 
Voigt profile fits to the lines used in the analysis are
shown in  Fig.~\ref{velp0002}.
\section{Results} 
\begin{figure*}
\centerline{\vbox{
\psfig{figure=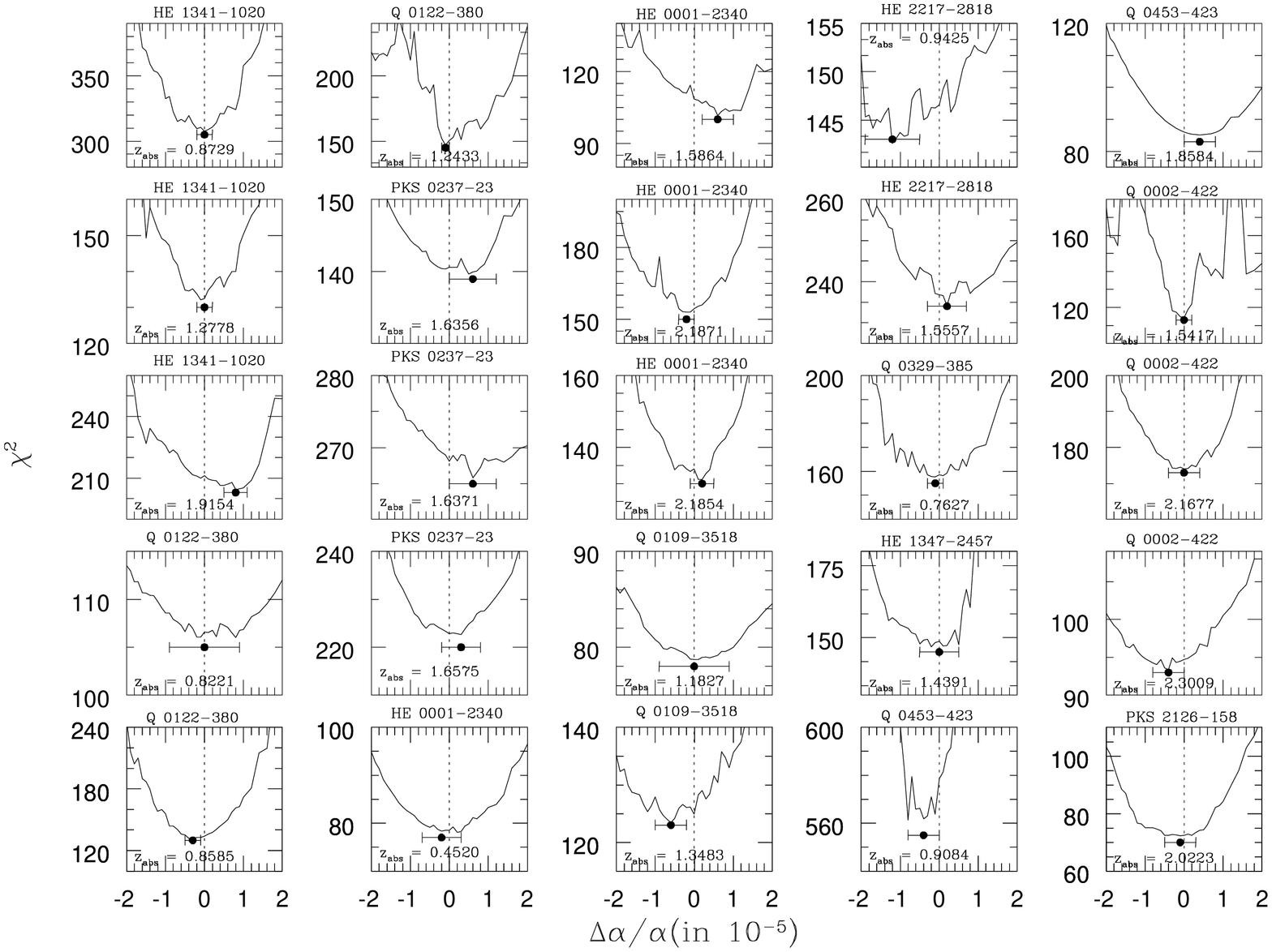,height=14.cm,width=18.cm,angle=270.}
%fig11.ps
}}
\caption[]{{\bf Results from the UVES sample:} $\chi^2$ is plotted
as a function of \dela for each system in our sample.  The minima
of the curve (marked with a dot) gives the best fitted value \dela.
The error in the measurement (error bar around the dot) is 
derived using $\Delta\chi^2 = 1$. Name of QSOs and 
redshifts \zabs are given in each panel. 
The vertical dotted line in each panel marks the 
location of \dela = 0. It is interesting to note that most of
the systems are consistent with \dela = 0 within error bars.
Two of the systems \zabs = 1.5864 toward HE0001$-$2340 and 
\zabs = 0.7627 toward Q~0329$-$385 are not considered in the
analysis as the only two components in these systems 
are blended as per our definision. 
}
\label{figchi}
\end{figure*}
\begin{figure*}
\centerline{\vbox{
\psfig{figure=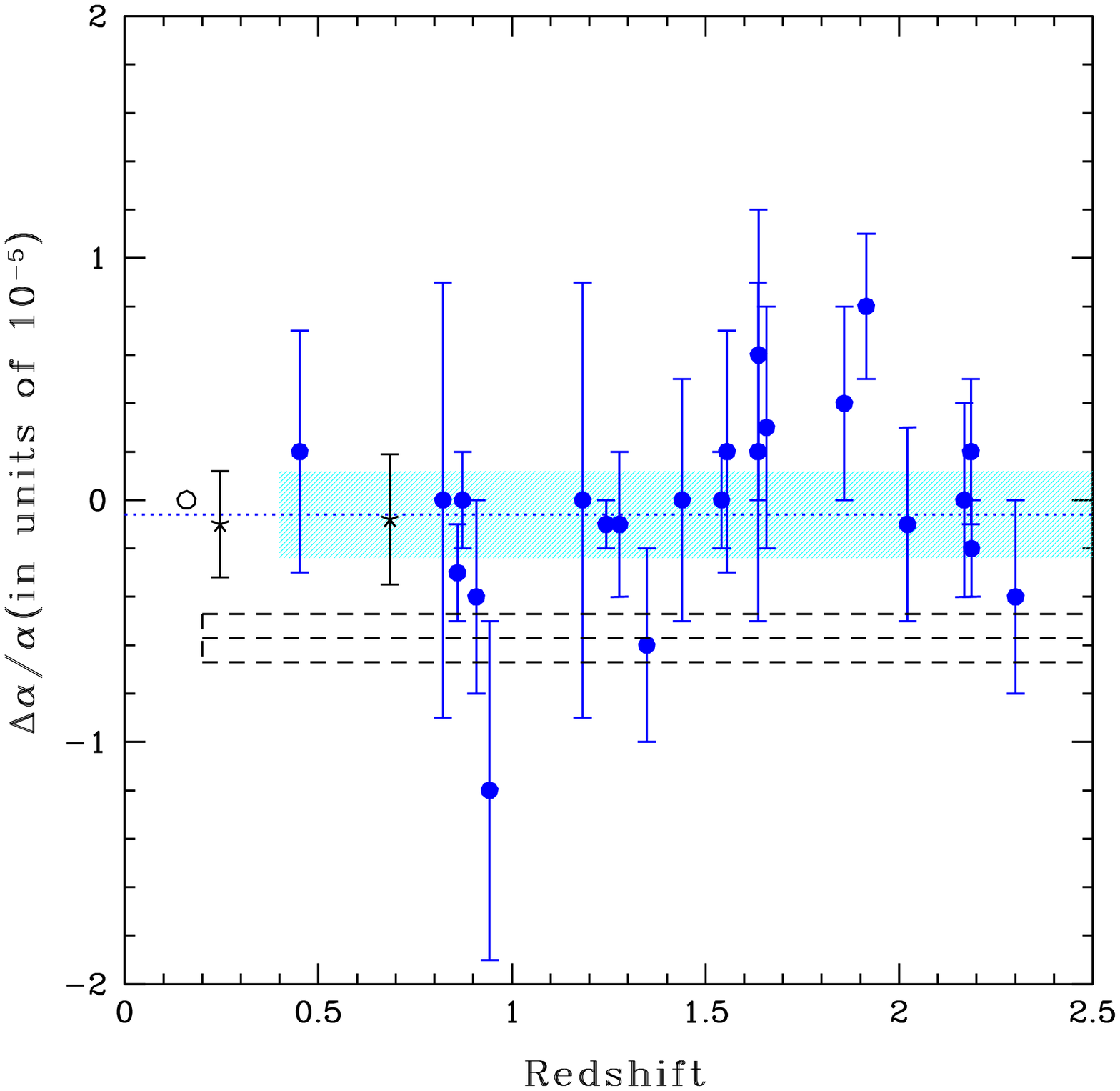,height=14.cm,width=18.cm,angle=0.}
%fig11.ps
}}
\caption[]{{\bf Results from the UVES sample:} 
The measured values of $\Delta\alpha/\alpha$ from
our sample (filled circles) are plotted
against the absorption redshifts of Mg~{\sc ii} systems. Each point is the 
best fitted value obtained for individual systems using $\chi^2$ 
minimization as demonstrated in Fig.~\ref{figchi}. 
The open circle and stars are the measurement from 
Oklo phenomenon and from molecular lines
respectively. The weighted mean and 1$\sigma$ range measured by Murphy et al.(2003)
are shown with the horizontal long dashed lines. Clearly most of our
measurements are inconsistent with this range.
The shadow region marks the weighted mean and its 3$\sigma$ error 
obtained from our study [${\rm <\Delta\alpha/\alpha>_w
 ~=~ (-0.06\pm0.06)\times 10^{-5}}$]. 
Our data gives a 3$\sigma$ constraint on the variation of \dela to be
$-2.5\times 10^{-16} ~{\rm yr}^{-1}\le(\Delta\alpha/\alpha\Delta t) 
\le 1.2\times 10^{-16}~{\rm yr}^{-1} $
in the case of flat universe with
$\Omega_\lambda = 0.7$, $\Omega_m$ = 0.3 and H$_0$ = 68 \kms Mpc$^{-1}$
for the median redshift of 1.55. 
}
\label{figfinal}
\end{figure*}
\subsection{Fitting procedure}
As discussed before the accuracy of the method depends on how well
absorption line profiles are modeled using Voigt profiles.
As can be seen from Table.~\ref{tabat} the relative shifts
between different Fe~{\sc ii} lines ({apart from Fe~{\sc ii}$\lambda1607$}) are small
compared to the shifts between the Fe~{\sc ii} lines and the anchors.
We use this fact to refine our fitting procedure.

We first fit all the Fe~{\sc ii} lines simultaneously using 
laboratory wavelengths ($\Delta\alpha/\alpha$~=~0).  
This allows us to find out about (i) bad pixels, (ii) unknown contaminations
and (iii) the velocity component structure in Fe~{\sc ii}. 
We therefore can remove inadequate profiles 
(because of contamination or bad pixels) from the \dela analysis.
Similar exercise was carried out for Mg~{\sc ii} 
doublets and other anchors.  
Based on these preliminary fits a first set of parameters is generated to
start the Voigt profile fitting procedure that includes \dela variations as described in  section 3.3. 

In all our fits we have tied up the redshifts of all species under consideration. 
We do not know a priori whether $b$ parameters of different species are identical 
or not. To take this uncertainty into account we perform the fit of the systems in
two cases:
(i) assuming that $b$ parameters of individual components are same for 
all species and  (ii) assuming different $b$ parameters for different
species. We determine the best fitted \dela values for both cases. It 
is found that in all cases the two values are consistent with one another 
within 1$\sigma$ uncertainty. For our final result we consider the fit that 
gives the smaller reduced $\chi^2$. 

\begin{table*}
\caption{Recovered value of \dela in individual systems}
{\tiny
\begin{tabular}{lccccccc}
\hline
\hline
\\
Name & $z_{\rm em}$  & \zabs&b& Transitions &N$_{\rm c}$&
\multicolumn{2}{c}{$\Delta\alpha/\alpha$ (in units of $10^{-5}$)} \\
     &               &      & &            & &case 1&case 2\\ 
\hline
\\
\\
HE1341$-$1020    &  2.135  &$0.872666\pm0.000046$&$5.65\pm0.13$&abdfgi&3&$+0.0\pm0.2(1.19)$&$-0.6\pm0.4(1.20)$\\
                 &         &$0.872841\pm0.000212$&$5.22\pm0.28$&abdfg& \\
                 &         &$0.872903\pm0.000195$&$3.28\pm0.28$&abdfg&\\
                 &         &$0.872755\pm0.000266$&$2.49\pm0.90$&ab&\\
                 &         &$0.872422\pm0.000089$&$2.59\pm0.30$&ab&\\
                 &         &$0.872477\pm0.000563$&$2.68$      &ab&\\
                 &         &$0.872568\pm0.000354$&$7.65\pm0.84$&ab&\\
                 &         &$0.873007\pm0.000206$&0.50         &ab& \\  
                 &         &$1.277834\pm0.000584$&$4.03\pm0.51$&abdefh&2&$+0.1\pm0.2(1.01)$&$+0.0\pm0.50(0.97)$\\
                 &         &$1.277780\pm0.000571$&$3.35\pm0.52$&abdefh&\\
                 &         &$1.915216\pm0.000628$&$2.57\pm0.73$&cdefhkl &4&$+0.8\pm0.3(1.49)$&$+0.6\pm0.3(1.48)$\\
                 &         &$1.915330\pm0.000226$&$3.40\pm0.23$&cdefhkl&\\
                 &         &$1.915463\pm0.000302$&$6.24\pm0.28$&cdefhkl&\\
                 &         &$1.915561\pm0.002269$&$3.86\pm1.44$&cdefhkl&\\
Q0122$-$380      &  2.190  &$0.822076\pm0.000032$&$5.69\pm0.07$&abdfghi&1&$+0.0\pm0.9(0.87)$&$+0.0\pm0.9(0.87)$ \\
                 &         &$0.822182\pm0.000457$&$6.41\pm1.28$&ab&\\
                 &         &$0.859270\pm0.000211$&$3.92\pm0.08$&abfghi&2&$-0.3\pm0.2(1.29)$&$-0.5\pm0.3(1.18)$\\
                 &         &$0.859318\pm0.000646$&$3.42\pm0.13$&abfghi&\\ 
                 &         &$1.243293\pm0.000109$&$5.09\pm0.12$&abdef&3&$-0.1\pm0.1(0.89)$&$-0.5\pm0.1(1.01)$\\
	         &         &$1.243374\pm0.000344$&$3.12\pm0.50$&abdef&\\
                 &         &$1.243483\pm0.000122$&$6.83\pm0.41$&abdef&\\
                 &         &$1.243589\pm0.000731$&$5.20\pm1.03$&ab&\\

PKS0237$-$23     &  2.222  &$1.635674\pm0.000708$&$9.23\pm1.06$&abk&2&$+0.2\pm0.7(0.82)^*$&$-0.2\pm0.3(0.86)^*$\\
                 &         &$1.635790\pm0.000132$&$2.56\pm0.34$&abdghk&\\
                 &         &$1.635883\pm0.000050$&$4.04\pm0.80$&abdghk&\\
                 &         &$1.636041\pm0.000152$&$4.20\pm0.33$&abk&\\
                 &         &$1.637128\pm0.000114$&$3.60\pm0.42$&abcdgh&5&$+0.6\pm0.6(1.16)^*$&$-0.4\pm1.3(1.19)^*$\\
                 &         &$1.636903\pm0.000119$&$0.40\pm0.13$&abdgh&\\
                 &         &$1.637331\pm0.000244$&$3.87\pm0.62$&abdgh&\\
                 &         &$1.637155\pm0.000209$&$15.4\pm0.41$&abcdgh&\\
                 &         &$1.637437\pm0.000170$&$6.17\pm0.24$&abdgh&\\
                 &         &$1.657390\pm0.000223$&$5.46\pm0.34$&dfjk&8&$+0.3\pm0.5(0.92)$&$+0.3\pm0.7(0.92)$\\
                 &         &$1.657172\pm0.000932$&$5.51\pm0.92$&dfjk&\\
                 &         &$1.657415\pm0.000348$&$11.57\pm0.34$&dfjk&\\
                 &         &$1.656864\pm0.001122$&$10.42\pm1.22$&fjk&\\
                 &         &$1.657066\pm0.001358$&$4.75\pm1.34$&fjk&\\
                 &         &$1.657723\pm0.000785$&$6.91\pm1.02$&fjk&\\
                 &         &$1.657884\pm0.000262$&$5.36\pm0.28$&dfjk&\\
                 &         &$1.657591\pm0.000622$&$2.16\pm0.17$&dfjk&\\
HE0001$-$2340    &  2.263  &$0.452060\pm0.000024$&$1.74\pm0.05$&abefhim&1&$+0.2\pm0.5(1.10)$&$-0.3\pm0.5(1.21)$\\
%
%                 &         &$0.949015\pm0.000298$&$5.56\pm0.39$&abfh&3&$-1.1\pm2.3(0.90)$&$-1.1\pm2.2(0.91)$\\
%                 &         &$0.949070\pm0.000515$&$4.41\pm0.52$&abfh&\\
%                 &         &$0.948917\pm0.000338$&$3.04\pm1.04$&abfh&\\
%
%                 &         &$1.586429\pm0.000066$&$1.58\pm0.15$ &efghij   &2&$+0.6\pm0.4(1.41)$& $+0.6\pm0.4(1.33)$\\
%                 &         &$1.586450\pm0.000060$&$6.37\pm0.77$&fhij   &\\
%
                 &         & $2.187149\pm0.001503$&$2.03\pm0.23$&bdegjk&3&$-0.2\pm0.2(1.20)$&$-0.2\pm0.3(1.31)$\\
                 &         & $2.187237\pm0.000129$&$4.73\pm1.06$&\\
                 &         & $2.187106\pm0.001420$&$6.61\pm0.56$&\\
                 &         & $2.185294\pm0.000089$&$2.29\pm0.19$&abdefjk&2&$+0.2\pm0.3(1.15)$&$-0.2\pm0.3(1.17)$\\
                 &         & $2.185452\pm0.000264$&$3.48\pm0.52$&abdefjk&\\
                 &         & $2.185256\pm0.000512$&$1.51$       &abjk&\\
                 &         & $2.185332\pm0.000839$&$1.77$       &abjk&\\
Q0109$-$3518     &  2.404  & $1.182683\pm0.000039$&$3.04\pm0.09$&abdfhi &1&$+0.0\pm0.8(0.98)$&$-0.5\pm0.5(0.98)$ \\
                 &         & $1.182770\pm0.000389$&$3.17\pm1.18$&ab  \\
                 &         & $1.348308\pm0.000204$&$2.28\pm0.21$&defghijk&2&$-0.6\pm0.4(1.08)$&$-0.6\pm0.4(1.08)$\\
           
                 &         &$1.348346\pm0.000415$&$3.72       $&defghijk&\\
%
%                &         & $1.350834\pm0.000124$&$1.61\pm0.14$&defghi&2&$+0.4\pm0.4$&$+0.4\pm0.4$\\
%                &         & $1.350782\pm0.000330$&$4.36\pm0.34$&defghi&\\
HE2217$-$2818    &  2.414  &$0.942330\pm0.000058$&$4.91\pm0.12$&abfh&5&$-1.2\pm0.7(0.90)^*$&$-1.6\pm0.8(1.36)$\\
                 &         &$0.942574\pm0.000067$&$4.79\pm0.15$&abfh& \\
                 &         &$0.942655\pm0.000119$&$3.27\pm0.75$&abfh& \\
                 &         &$0.942701\pm0.002092$&$8.88\pm2.20$&abfh&\\
                 &         &$0.942435\pm0.000454$&$7.38\pm0.64$&abfh&\\
                 &         &$0.942476\pm0.000282$&$4.82$       &abfh&\\
                 &         &$1.555425\pm0.000029$&$2.52\pm0.06$&ab&1&$+0.2\pm0.5(1.22)^*$&$+0.0\pm0.3(1.35)^*$\\
                 &         &$1.555580\pm0.000098$&$2.35\pm0.06$&ab&\\
                 &         &$1.555721\pm0.000178$&$3.08\pm0.39$&ab&\\
                 &         &$1.555886\pm0.000041$&$4.98\pm0.07$&abdfghk&\\
                 &         &$1.556162\pm0.000055$&$3.76\pm0.11$&ab&\\
%
%Q0329$-$385      &  2.435  &$0.762727\pm0.000066$&$3.45\pm0.15$&abefghi&2&$-0.1\pm0.2(1.45)$&$-0.6\pm0.4(1.44)$\\
%                 &         &$0.762748\pm0.000252$&$3.23\pm0.19$&\\
%
HE1347$-$2457    &  2.611  &$1.439541$&40.91&k&3&$-0.0\pm0.5(1.10)$&$-0.4\pm0.5(1.11)$\\
                 &         &$1.439219\pm0.002721$&$14.2\pm0.90$&dgkl&\\
                 &         &$1.439243\pm0.000365$&$6.96\pm0.36$&dgkl&\\
                 &         &$1.439349\pm0.000195$&$2.61\pm0.07$&dgkl&\\
                 &         &$1.438934$ &$ 5.97$&k\\
Q0453$-$423      &  2.658  &$0.908131\pm0.000028$&$2.91\pm0.10$&abdfh&8&$-0.4\pm0.4(1.82)$&$-0.4\pm0.3(1.93)$\\
                 &         &$0.908225\pm0.000306$&$3.62\pm0.76$&abdfh&\\
                 &         &$0.908402\pm0.000044$&$3.90\pm0.10$&abdfh&\\
                 &         &$0.908490\pm0.000122$&$7.43\pm0.33$&abdfh&\\
                 &         &$0.908588\pm0.000069$&$2.78\pm0.15$&abdfh&\\
                 &         &$0.908634\pm0.000107$&$16.31\pm0.22$&abdfh&\\
                 &         &$0.908771\pm0.000122$&$1.42\pm0.32$&abdfh&\\
                 &         &$0.908266           $ &$22.97\pm2.58$&ab&\\

                 &         &$1.858364\pm0.000078$&$6.13\pm0.11$&abdfh&1&$+0.4\pm0.4(1.13)$&$+0.2\pm0.5(1.16)$\\
%
%PKS0329$-$255    &  2.703  & \\
Q0002$-$422      &  2.767  &$1.541474\pm0.000194$&$7.61\pm0.34$&ab&3&$+0.0\pm0.2(0.66)$&$-0.3\pm0.2(0.65)$\\
                 &         &$1.541799\pm0.000687$&$11.86\pm0.70$&abdfgh& &\\
                 &         &$1.541869\pm0.000100$&$4.09\pm0.18$&abdfgh     & &\\
                 &         &$1.541935\pm0.000267$&$5.28\pm0.24$&abdfgh     & &\\
                 &         &$1.541605\pm0.000397$&$4.55\pm0.91$&ab     & &\\

                 &         &$2.167849\pm0.000336$&$3.42\pm0.28$&cdfghjk      &2&$+0.0\pm0.4(1.03)^*$&$-0.7\pm0.4(1.03)^*$\\
                 &         &$2.167783\pm0.000529$&$2.568$      &cdfghjk      &\\   
                 &         &$2.300832\pm0.000231$&$4.78\pm0.19$& cdjk     &3&$-0.4\pm0.4(0.99)$&$-0.4\pm0.4(1.13)$\\
                 &         &$2.300997\pm0.000198$&$5.61\pm0.15$& dfjk     &\\
                 &         &$2.300619\pm0.000111$&$2.94\pm0.11$&dfjk\\
%HE0151$-$4326    &  2.789  & \\

%
%HE2347$-$4342    &  2.871  &$1.796113\pm0.000445$&$3.14\pm0.86$&abk&2&$-0.4\pm2.0(1.01)$&.... \\
%                 &         &$1.796277\pm0.000939$&$2.08\pm0.63$&abfhk&\\
%                 &         &$1.796219\pm0.000759$&$3.03\pm0.55$&abfhk&\\
%
%HE0940$-$1050    &  3.084  & \\
%
PKS2126$-$158    &  3.280  &$2.021923\pm0.000075$&$4.87\pm0.10$&bdfgk&1&$-0.1\pm0.4(1.19)$&$-0.9\pm0.5(1.21)$\\
                 &         &$2.022001\pm0.000612$&$2.54$        &bk&\\
\hline
\hline
\end{tabular}
\label{tabfinal}
}
\end{table*}
\begin{table}
\caption{Results for the weak systems that show
at least two detectable Fe~{\sc ii} lines:}
\begin{tabular}{llc}
\hline
\multicolumn {1}{c}{QSO}&\zabs&\dela (in 10$^{-5}$)\\
\hline
\hline
Q~$0122-380$    &1.9102 &$-0.9\pm1.6$\\
PKS~$1448-232$  &1.5847 &$+0.4\pm2.0$\\
PKS~$0237-23$   &1.1846 &$-0.9\pm2.7$\\
HE~$0001-234$   &0.9489 &$-1.1\pm2.3$\\
HE~$2217-2818$  &1.6277 &$+0.3\pm0.6$\\
Q~$0329-385$    &1.4379 &$-0.4\pm0.6$\\
HE~$1347-2457$  &1.5082 &$-3.9\pm1.4$\\
PKS~$0329-255$  &0.9926 &$-0.3\pm1.3$\\
HE~$2347-4342$  &1.7962 &$-0.4\pm2.0$\\
\hline
\end{tabular}
\label{tabweak}
\end{table}

\subsection{Determination of \dela}

The $\chi^2$ analysis for individual systems is shown in
Fig.~\ref{figchi}. Here, we plot $\chi^2$ as a function of 
\dela. The dot with error-bar in each panel gives our best fit estimate of 
\dela and its errors. The vertical lines in each panel shows the value
\dela = 0. It is clear from the figure that apart from few cases
the curve is well behaved and the best fitted values are consistent
with zero in most cases. The numerical values of the fits are 
given in Table.~\ref{tabfinal}. In this table we list
in the first four columns the name of the QSO, the emission 
redshift (\zem), absorption redshifts (\zabs) 
and velocity dispersions $b$ of individual components. When
the best fit solutions are obtained with different $b$ parameters 
the $b$ values quoted are for the main anchor line (Mg~{\sc ii}, Si~{\sc ii}
or Al~{\sc ii} lines). Column \#5 lists the transitions used in
the analysis with the same notations as in Table~\ref{tabat}.
$N_{\rm c}$ given in column \#6 is the number of subcomponents used 
to fit the Fe~{\sc ii} absorption profiles. 
The last two columns give the determination of \dela obtained using
laboratory wavelengths of Mg~{\sc ii}, Mg~{\sc i}, and Si~{\sc ii}
absorption lines with terrestrial isotopic abundances (case 1) and 
with the wavelengths of the dominant isotopes (case 2, see next Section). 
Numbers in brackets are the reduced $\chi^2$ for the fit. The numbers 
marked with a asterisks ($^*$) are values obtained using different $b$ values
for different species. 

In Fig.~\ref{figfinal} we plot individual determinations of \dela 
as a function of \zabs. We also plot the exisiting results from
the literature.  
The horizontal dotted line gives the
weighted mean of our sample with 1/error$^2$ weighting.  We estimate
the error in the weighted mean using the standard equation,
\begin{equation}
{\rm
Error ~in}~x_{\rm w} ~=~ \sqrt{{{\rm \Sigma_i^N} w_{\rm i}(x_{\rm i}-x_{\rm w})^2}\over 
{({\rm N}-1){\rm \Sigma_i^N} w_{\rm i} }}.
%}
\end{equation}

Here, $x_{\rm w}$ is the weighted mean of the variable $x_{\rm i}$ of 
sample size N whose weights are $w_{\rm i}$. 
The shaded region passing through most of the error bars is 
our measured  weighted mean
and its 3$\sigma$ error.
The histogram in the left
hand side of the panel shows the distribution of $\Delta\alpha$/$\alpha$.
The weighted mean with 1/$\sigma^2$ weighting obtained for our
sample is $-0.06\pm0.06\times10^{-5}$ and the
standard deviation
in our measurements around the mean is 0.41$\times10^{-5}$. All the
points used are consistent with this weighed mean value with
a reduced $\chi^2$ of 0.95. As can be seen from  the figure
there are two points that deviate by more than 1$\sigma$ from
derived weighted mean value. These are from the \zabs = 0.9425 system toward
HE~2217$-2818$ and the \zabs = 1.9154 system toward HE~1341$-$1020
(see also the $\chi^2$ curve for these systems in  Fig.~\ref{figchi}).
As can be seen from Fig.~\ref{velp0109}, the Fe~{\sc ii} lines
in the \zabs = 0.9425 system are weak (barely above the cutoff). The reduced 
$\chi^2$ is dominated by uncertainties in fitting the Mg~{\sc ii}
profiles.  
In the case of the latter system the effect could just
be due to the odd pixel in the Si~{\sc ii}$\lambda$1808 close
to $v$ = 0 \kms (see Fig.~\ref{velp1}).

\begin{table*}
\caption{Summary of results for various sub-samples:}
\centerline{
\begin{tabular}{lcccccc}
\hline
\hline
\multicolumn {1}{c}{Sample}& Number of & z &
\multicolumn{3}{c}{\dela}&$\chi^2_{\rm W}$\\
                           & systems   &   & mean & weighted mean&$\sigma$&\\
\hline
Single+double(case 1) & 12 & 1.54 &$+0.01\pm0.15$&$-0.08\pm0.07$&0.27&0.55\\
Full sample(case 1)   & 23 & 1.54 &$-0.02\pm0.10$&$-0.06\pm0.06$&0.41&0.95\\
Weak(case 1)          &  9 & 1.51 &$-0.80\pm0.58$&$-0.40\pm0.36$&1.27&1.00\\
Weak+All(case 1)      & 32 & 1.51 &$-0.24\pm0.18$&$-0.07\pm0.06$&0.81&0.96\\
Single+double(case 2) & 12 & 1.54 &$-0.31\pm0.14$&$-0.33\pm0.09$&0.33&0.63\\
Full Sample(case 2)   & 23 & 1.54 &$-0.33\pm0.11$&$-0.36\pm0.06$&0.44&1.06\\
 \hline
\multicolumn{7}{l}{Case 1: Laboratory wavelengths given in
Table.~\ref{tabat} are used.}\\
\multicolumn{7}{l}{ Case 2: Rest wavelengths of dominant
isotopes for Mg and Si are used.}
\end{tabular}
}
\label{tabsubsample}
\end{table*}

Just for completeness we also fitted the weak systems
that show at least two Fe~{\sc ii} lines. 
The recovered values of \dela based on these
are given in Table~\ref{tabweak}. 
As expected the
individual measurements have large errors.The weighted mean
value of \dela measured from these systems is
$-0.40\pm0.36$.  Most of the individual measurements and the
weighted mean
are consistent with the value obtained from our main sample.

A summary of the results for different sub-samples are 
given in Table~\ref{tabsubsample}. The sample identification
is given in the first column. Number of systems used and 
the median redshift of the sample are given in column
\#2 and \#3. The mean, weighted mean and $\sigma$ of 
the measured \dela values are given in columns \#4, \#5 and \#6 
respectively. Last column gives the reduced $\chi^2_{\rm W}$
obtained from all our measurements 
for the measured weighted mean. 
It can be seen that in none of our subsamples we find a significant
change in $\alpha$.

\subsection{The magnesium isotopic abundance}

One major uncertainty in the many-multiplet analysis comes from the 
determination of the effective rest wavelengths. Even though laboratory
wavelengths are measured with a precision of a few 0.1 m\AA, values given 
in Table.~\ref{tabat} assume terrestrial isotopic 
abundances. This assumption may not be valid at high redshift.
In the Astrophysical settings the effect of isotopic shifts
could be important for Si and Mg (see Murphy et al. 2003, Ashenfelter 
et al. 2003). Indeed, the relative abundances
of different isotopes may depend on the overall metallicity of
the gas. Gay \& Lambert (2000) have shown that the abundance of $^{25}$Mg and 
$^{26}$Mg relative to $^{24}$Mg decreases with decreasing 
metallicity. In the low metallicity gas ($\rm Z\le0.01Z_\odot$) 
most of the metals
will be in their dominant isotopic state. Thus in the extreme
case of very low metallicity, the effective
rest wavelengths could take values in the range from
terrestrial composite wavelength to wavelengths 
corresponding to the dominant isotope.
This range is less than 0.5 m\AA~ for 
Mg~{\sc i} and Si~{\sc ii} absorption lines
but is of the order of 1~m\AA~ for Mg~{\sc ii} lines.
In order to accommodate this uncertainty we fit the systems 
using the wavelengths of the species from the dominant isotope.
The measurements are given in the last column of Table~3. 
As expected, using these abundances leads to 
a lower $\alpha$ determination (see Table~5). Note however 
that even in this extreme case, the variation stays smaller
than what has been claimed from previous studies.
Note that the assumption of very low metallicity is extreme as:
(i) the systems in the z range 
0.4$-$2 are more likely to have metallicity
larger than 0.1~Z$_{\odot}$ (e.g. Ledoux et al. 2002);
(ii) the measured mean ratios of Mg$^{24}$:Mg$^{25}$:Mg$^{26}$
in the cool dwarf with metallicity Z = $-1.5$ to $-1.0$ Z$_\odot$
is 80:10:10 (from Table. 1 of Yong et al. 2003). This gives
the weighted mean wavelengths close to the terrestrial 
wavelengths,  and (iii) We also notice that the minimum $\chi^2$ 
for the fit in most cases (apart from 4 cases) are better 
when we use the laboratory wavelengths in our analysis.
Therefore, although some additional uncertainty and scatter
could come from the isotopic abundances being different
from that of terrestrial composition, 
our result using the laboratory wavelengths is
most probably robust. 
\section{Conclusion}
We have applied the MM method to a homogeneous 
sample of 50 Mg~{\sc ii} systems observed along 18 QSO lines of sight
observed with UVES at the VLT.  Using extensive simulations
we show our
Voigt profile fitting procedure 
works well to recover the 
input value of \dela for simple single component systems. 
We can recover the input \dela values with an error of $\sigma = 
0.23\times10^{-5}$
in the case of individual components
for data of quality comparable to the UVES data (S/N~$\sim$~70). 
We show that the uncertainty is about twice larger for data of S/N~$\sim$~30 as
used by previous surveys. In addition at these S/N ratio, the use of strongly blended 
systems leads to larger uncertainties as well. We show that weak lines should
be avoided in the determination as their use can lead to false determination
of \dela.

We device the selection criteria for our sample using extensive simulations.
This leads us to 23 systems avoiding stongly blended systems and 
systems with weak absorption lines.
Individual systems and fits to all absorption lines 
are discussed in detail. We also avoided the two sub-DLAs from the 
analysis. In one of the systems that is in 
common with Murphy 
et al. (2003) our results are consistent with the earlier measurement
albeit with smaller error (see section 5.3.1).

The weighted mean value of the variation in the fine-structure
constant obtained from our analysis over the redshift 
range ${ 0.4\le z\le 2.3}$ is 
${\Delta\alpha/\alpha}$~=~${-0.06\pm0.06\times10^{-5}}$. 
For the median redshift of our sample (z = 1.54)
we obtain a 3$\sigma$ constraint on the variation of \dela to be
$-2.5\times 10^{-16} ~{\rm yr}^{-1}\le(\Delta\alpha/\alpha\Delta t) 
\le 1.2\times 10^{-16}~{\rm yr}^{-1} $
in the case of flat universe with
$\Omega_\lambda = 0.7$, $\Omega_m$ = 0.3 and H$_0$ = 68 \kms Mpc$^{-1}$. 
To our knowledge this
is the strongest constraint from QSO absorption line
studies till date. 

We show the effect of varying the isotopic abundances.
Only if we assume isotopic abundances for very low metallicity 
would $\alpha$ show some variation. This is non acceptable 
for (i) assuming that all systems in the redshift range
0.4$-$2 have extremely low metallicity is incorrect as metals are
seen in these systems and metallicity is measured to be
larger than 0.1~Z$_{\odot}$ (e.g. Ledoux et al. 2002);
(ii) at Z~$\sim$~0.1~Z$_{\odot}$, which would be a more
reasonable assumption, the composite wavelength we get
will be close to the laboratory value we use in our
study 
(we use Yong et al. 2003 for isotopic abundances).  
Therefore, although some additional uncertainty and scatter
could come from the isotopic abundances being different
from terrestrial, our result using laboratory wavelengths  is
most probably robust. 

In conclusion, our study does not support claims by previous authors
of a statistically significant change in \dela with 
cosmic time at $z$~$>$~0.5. 
Our result still does allow smaller variations in 
excess of what is found based on the Oklo phenomenon. Future very 
high resolution (R$\sim$100,000) spectroscopic
studies are needed to probe the variations in $\alpha$ with much 
better accuracy.

\section*{Acknowledgments}
This work is based on observations collected during programme 166.A-0106
(PI: Jacqueline Bergeron) of the European Southern Observatory with the
Ultra-violet and Visible Echelle
Spectrograph mounted on the 8.2~m KUEYEN telescope operated at the Paranal
Observatory, Chile. PPJ thanks E. Vangioni-Flam and J. P. Uzan for
fruitful discussions.
HC thanks CSIR, INDIA for the grant award
No. 9/545(18)/2KI/EMR-I. RS thanks CNRS/IAP
for the hospitality.
We gratefully acknowledge support from the Indo-French
Centre for the Promotion of Advanced Research (Centre Franco-Indien pour
la Promotion de la Recherche Avanc\'ee) under contract No. 3004-A.


\begin{thebibliography}{}
% T. Ashenfelter, Grant J. Mathews, Keith A. Olive
\bibitem {}  Ashenfelter, T., Mathews, G. J., Olive, K. A., 2003,
astro-ph/0309197.
%\bibitem{} Barlow, R. J. 2002, ``STATISTICS: A guide to the use of
%statistical methods in the physical sciences'', Wiley, p106-107
%
\bibitem{} Bahcall, J. N., Sargent, W. L. W., \& Schmidt, M. 1967, 
ApJ, 149, L11
%
\bibitem{} Bahcall, J. N., Steinhardt, C. L., \& Schlegel, D. 2003, 
astro-ph/0301057
%
\bibitem{} Bergeson S. D. et al. 1996, ApJ, 464, 1044
%
\bibitem{} Chengalur, J. N., Kanekar, N., 2003, /astro-ph/0310764
%
\bibitem{} Cowie, L. L., \& Songaila, A., 1995, ApJ, 453, 596
%
\bibitem{} Dzuba, V. A., et. al. 2002, Phys. Rev A., 66, 022501
%
\bibitem{} Dzuba, V.A., Flambaum, V. V., \& Webb, J. K., 1999, PRL, 82, 888
%
\bibitem{} Edl\'en, B,1966, {\sl Metrologia}, {\bf 2},71.
%
\bibitem{} Gay, P., \& Lambert, F. L. 2000, ApJ, 533, 260
%
\bibitem{} Griesmann U., \& Kling, R., 2000, ApJ, 536, L113
%
\bibitem{} Khare, P., Srianand, R., York, D. G.,
 Green, R., Welty, D., Huang,K., \& Bechtold, J.  1997, MNRAS, 285, 167
%
\bibitem{} Ledoux, C., Bergeron, J., \& Petitjean, P. 2002, A\&A, 385, 802
%
\bibitem{} Levshakov, S. A. 1994, MNRAS, 269, 339
%
\bibitem{} Mohr, P. J., \& Taylor, B. N. 2000, Rev. Mod. Phys., 72, 351
%
\bibitem{} Morton, D. C., 1991, ApJS, 77, 119 
%
\bibitem{} Morton, D. C., 1992, ApJS, 81, 883 
%
\bibitem{} Murphy, M. T., Webb, J., Flambaum, V., Prochaska, J. X., 
\& Wolfe, A. M. 2001, MNRAS, 327, 1237
%
\bibitem{} Murphy, M. T., Webb, J., Flambaum, V., Drinkwater, M. J.,
Combes, F., Wiklind, T. 2001a, MNRAS, 327, 1244
%
\bibitem{}  Murphy, M. T., Webb, J. K., Flambaum, V. V.,
2003, MNRAS, 345, 609
%
\bibitem{} Nave, G., et al., 1991, J. Opt. Soc. Am. B, 8, 2028
%
\bibitem{} Petitjean, P., \& Aracil, B., 2003, A\&A, submitted.
%
\bibitem{} Pickering, J. C., et al., 2002, MNRAS,319,163
%
\bibitem{} Pickering, J. C., et al., 1998, MNRAS,300,131
%
\bibitem{} Potekhin, A. Y., \& Varshalovich, D. A. 1994, A\&AS, 104, 89
%
\bibitem{} Press, W., et, al., 2000, ``Numerical Recipes in
          Fortran:The art of Scientific Computing'' Foundation Books,p690-691
%
\bibitem{} Prochaska J. X. et al., 2001, ApJS, 137, 21
%
%\bibitem{} Savedoff, M. P. 1956, Nature, 178, 688
%
\bibitem{} Srianand, R., \& Khare, P. 1993, ApJ, 413, 486
%
\bibitem{} Stumpff, P., 1980, A\&ASS, 41,1
%
\bibitem{} Uzan,J., 2003, RvMP,75, 403
\bibitem{} Varshalovich, D. A., Panchuk, V. E., \& Ivanchik, A. V.
1996, Astron. Lett., 22, 6.
%
\bibitem{}  Webb, J. K., et al. 2001, PRL, {87}, 091301
%
\bibitem{} Wolfe, A. M., Brown, R. L., \& Roberts, M. S. 1976, Phys. Rev. 
Lett., 37, 177
%
\bibitem{} Yong, D., Grundahl, F., Lambert, D. L., Nissen, P. E., \& 
Shetrone, M. D. 2003, A\&A, 402, 985
%
\end{thebibliography}
\end{document}